\documentclass[12pt]{article}

\usepackage{paper2e}
\usepackage{mydefs2e}
\usepackage{xspace}

\usepackage{boxedeps}
\HideDisplacementBoxes
\SetepsfEPSFSpecial 

\renewcommand{\Box}{\,\raisebox{-.45pt}{\drawsquare{7}{0.6}}\,}
\renewcommand{\c}{{\rm c}}

\newcommand{\MP}{M_{\rm P}}
\renewcommand{\d}{\partial}

\newcommand{\half}{1\hskip -0.3pt/2}

\renewcommand{\bex}[1][]{%  optional argument is label of exercise
    \stepcounter{exercise}%
    \begin{boxedtext}%
    \noindent{\bfseries\theexercisename\
\arabic{exercise}\ifempty#1\else\ #1\fi:\ }}

\begin{document}

% ---------------------------------------------------------------------
% Title page
% ---------------------------------------------------------------------
\begin{titlepage}
\preprint{}

\title{2004 TASI Lectures on\\\medskip
Supersymmetry Breaking}

\author{Markus A. Luty}

\address{Physics Department
\\University of Maryland\\
College Park, MD 20742}% $\,$\footnote{Permanent address}}

%\address{Jefferson Laboratory of Physics, Harvard University,
%Cambridge, MA 02138}

%\address{Physics Department, Boston University,
%Boston, MA 02215}

\address{\tt luty@umd.edu}

\begin{abstract}
These lectures give an introduction to the problem of finding a realistic
and natural extension of the standard model based on spontaneously broken
supersymmetry.
Topics discussed at some length include the effective field theory paradigm,
coupling constants as superfield spurions,
gauge mediated supersymmetry breaking,
and anomaly mediated supersymmetry breaking,
including an extensive introduction to 
supergravity relevant for phenomenology.
\end{abstract}

\end{titlepage}

% ------------------------------------------------------------------
% Table of contents
% ------------------------------------------------------------------

\pagenumbering{roman}
\tableofcontents
\newpage
\setcounter{page}{1}%  reset page #
\pagenumbering{arabic}

% ------------------------------------------------------------------
\section{Introduction}
% ------------------------------------------------------------------
Our present understanding of particle physics is based on
effective quantum field theory.
Quantum field theory is the inevitable result of combining
quantum mechanics and special relativity, the two great scientific
revolutions of the early twentieth century.
An \emph{effective} quantum field theory is one that includes only the
degrees of freedom that are kinematically accessible in a particular
class of experiments.
Presently, the highest energies probed in accelerator experiments
are in the 100~GeV range, and the standard model is an
effective quantum field theory that
describes all physical phenomena at energies of order 100~GeV and below.

The standard model contains a Higgs scalar that
has not been observed as of this writing.
Therefore, the minimal effective theory that describes the present data
does not contain a Higgs scalar.
This effective theory allows for the possibility that the dynamics that breaks
electroweak symmetry does not involve elementary scalars (as in
`technicolor' theories, for example).
This effective theory necessarily breaks down at energies of order
a TeV, and therefore new physics must appear below a TeV.
This is precisely the energy range that will be explored by the
LHC starting in 2007-2008, which is therefore all but guaranteed to discover
the interactions that give rise to electroweak symmetry breaking,
whether or not it involves a Higgs boson.

If the physics that breaks electroweak symmetry does not involve
particles with masses below a TeV, then it must be strongly coupled at a TeV.
Contrary to what is sometimes stated, 
precision electroweak experiments have not ruled
out theories of this kind.
For example, if we estimate the size of the $S$ and $T$ parameters
assuming that the electroweak symmetry breaking sector is strongly
coupled at a TeV with no large or small parameters, we obtain
\beq
\De S \sim \frac{1}{\pi},
\qquad
\De T \sim \frac{1}{4\pi},
\eeq
which are near the current experimental limits.
The strongly-coupled models that \emph{are} ruled out are those
that contain $N \gg 1$ degrees of freedom at the TeV scale;
in these models the estimates for $S$ and $T$ above are multiplied by $N$.
Another difficulty with building models of strongly-coupled electroweak
symmetry breaking is incorporating quark mixing without large flavor-changing
neutral current effects.
However, given our profound ignorance of strongly-coupled quantum field theory,
it may be prudent to keep an open mind.
Fortunately, the LHC will tell us the answer soon.

The subject of these lectures is weakly-coupled supersymmetry (`SUSY').
As you have heard in previous lectures,
there are several hints that SUSY is correct.
First, the simplest supersymmetric grand unified models predict a
precise relation among the three gauge couplings of the standard model
that is in excellent agreement with observation.
Second, the best fit to precision electroweak data is obtained with a
Higgs boson with a mass close to the experimental lower limit
of $114\GeV$ from LEP.
Such a light Higgs boson arises automatically in SUSY.
Finally, SUSY naturally contains a viable cold dark matter candidate.

If SUSY is in fact discovered at the LHC, it will be the culmination of decades
of work by many hands, starting with general theoretical investigations
of spacetime
symmetries and the construction of supersymmetric quantum field theories,
to the realization that SUSY can solve the hierarchy problem
and the construction of realistic models of broken supersymmetry.
It will be an intellectual triumph comparable to general relativity,
another physical theory that fundamentally changed our view
of space and time, and which was also proposed based on very general
theoretical considerations and later spectacularly verified by experiment.

If SUSY is realized in nature, it must be broken.
In this case, the pattern of SUSY breaking can give us a great deal of
information about physics at much higher energy scales.
There are a number of theoretically well-motivated mechanisms for
SUSY breaking, each of which give distinct patterns of SUSY breaking
that can be experimentally probed at the weak scale.
If nature is supersymmetric at the weak scale, then the experimental
program in particle physics after the LHC turns on
will be largely the study of superpartners.

These are exciting prospects, but we are not there yet.
There is at present no direct experimental evidence for superpartners,
or a light Higgs boson.
Indirect experimental constraints place strong constraints on the
simplest SUSY models, requiring either accidental cancelations 
(fine tuning) or
additional non-minimal structure in the theory.
It is premature to say that these constraints rule out the idea
of SUSY, but they must be addressed by any serious proposal 
for weak scale SUSY.
%I think it is fair to say that there is at present no fully
%satisfactory model of supersymmetry breaking, which to me means a
%model with no fine-tuned parameters, that naturally incorporates
%the successes of SUSY, and that is simple enough to have the ring
%of truth.

These lectures will review both the progress that has been made in
constructing realistic models of SUSY breaking, and the problems
faced by SUSY in general, and these models in particular.
It is my hope that these lectures will challenge and
inspire---rather than discourage---the next generation of particle
physicists.

% ------------------------------------------------------------------
\section{Effective Field Theory and Naturalness}
% ------------------------------------------------------------------
If we take seriously the idea that the standard model is an
effective field theory, then the coupling constants of the
standard model are not to be viewed as fundamental parameters.
Rather they are to be thought of as effective couplings determined by
a more fundamental theory.
How do we know there is a more fundamental theory?
For one thing,
we hope that there is a more fundamental theory
that explains the $\sim 20$ free parameters of the standard model
(mostly masses and mixings).
One piece of evidence for a simpler fundamental theory comes
from the fact that the standard model gauge couplings approximately
unify at a scale $M_{\rm GUT} \sim 10^{16}\GeV$.
This is evidence that $M_{\rm GUT}$ is a scale of new physics
described by a more fundamental theory.
Another evidence for a new scale is the fact that gravity becomes
strongly interacting at the scale $\MP \sim 10^{19}\GeV$, and
we expect new physics at that scale.
Finally, as we review below, the recent
observation of neutrino mixing suggests
the existence of another scale in physics of order $10^{15}\GeV$.
These scales are so large that we cannot ever hope to probe them directly
in accelerator experiments.
The best that we can do is to understand how the effective couplings
that we can measure are determined by the more fundamental theory.

%A quantum field theory is determined by a Lagrangian, which can in
%principle contain an infinite number of terms.
%If we view the theory as an effective theory derived from a more
%fundamental theory at a scale $\La$, then we expect the dimensionful
%parameters of the theory to be controlled by the scale $\La$.

% ------------------------------------------------------------------
\subsection{Matching in a Toy Model}
Let us consider a simple example that shows how effective couplings
are determined from an underlying theory.
Consider a renormalizable theory consisting of a
real scalar $h$ coupled to a Dirac fermion field $\psi$:
\beq[Lfund]
\scr{L} = \bar\psi i \sla\d \psi
+ \sfrac 12 (\d h)^2 - \sfrac 12 M^2 h^2
- \frac{\la}{4!} h^4 + y h \bar\psi \psi.
\eeq
This theory has a discrete chiral symmetry
\beq[discsymm]
\psi \mapsto \ga_5 \psi,
\qquad
h \mapsto -h,
\eeq
that forbids a fermion mass term, since
$\bar\psi \psi \mapsto -\bar\psi \psi$.
For processes with energy $E \ll M$, the scalar is too heavy
to be produced.
We should therefore be able to describe these processes using an
effective theory containing only the fermions.

We determine the effective theory by matching to the fundamental
theory.
Let us consider fermion scattering.
In the fundamental theory, we have at tree level
\beq
\BoxedEPSF{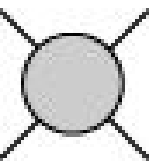} =
\BoxedEPSF{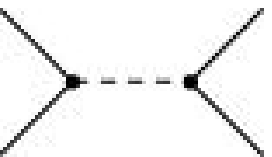} + \hbox{\rm crossed},
\eeq
while in the effective theory the scattering comes from an
effective 4-fermion coupling:
\beq
\BoxedEPSF{blob_small.eps} = \BoxedEPSF{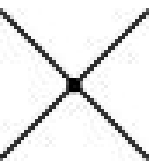}.
\eeq
Matching is simply demanding that these two expressions agree
order by order in an expansion in $1/M$.
This corresponds to expanding the scalar propagator in inverse
powers of the large mass:
\beq
\frac{1}{p^2 - M^2} = -\frac{1}{M^2} - \frac{p^2}{M^4}
+ \scr{O}(p^4 /M^6).
\eeq
Equivalently, we can solve the classical equations of motion
for $h$ in the fundamental theory order by order in $1/M^2$:
\beq
h &= \frac{1}{M^2} \left[
y \bar\psi \psi - \Box h - \frac{\la}{3!} h^3 \right]
\\
&= \frac{y}{M^2} \bar\psi \psi - \frac{y}{M^4} \Box (\bar\psi \psi)
+ \scr{O}(1/M^6).
\eeq
Substituting this into the lagrangain, we obtain the effective Lagrangian
at tree level
\beq
\scr{L}_{\rm eff} = \bar\psi i \sla\d \psi
+ \frac{y^2}{2 M^2} (\bar\psi \psi)^2
- \frac{y^2}{2 M^4} \bar\psi \psi \Box (\bar\psi \psi)
+ \scr{O}(1/M^6).
\eeq
This effective Lagrangian will give the same results as the Lagrangian
\Eq{Lfund} for all tree level processes, up to corrections of order $1/M^6$.

We can continue the matching procedure to include loop corrections,
as shown in Fig.~2.
The new feature that arises here is the presence of UV divergences
in both the fundamental and effective theories.
However, the matching ensures that the results in the effective
theory are finite, and this determines the counterterms that
are required in the effective theory.
The result is that the renormalized couplings in the effective
theory are a power series in $1/M$ even at loop level.

%To make this well-defined, we must choose a renormalization procedure
%for both the fundamental and the effective theory, but once this is
%done we obtain an effective Lagrangian that reproduces the physics
%of the fundamental Lagrangian at loop level.
%%The effective theory contains UV divergences that are not present
%%in the fundamental theory, but the matching to the effective theory
%%automatically generates counterterms that cancel these divergences.
\bex
Carry out the one-loop matching of the four fermion vertex
using a momentum space cutoff regulator in both
the fundamental and the effective theory.
If we write the effective Lagrangian as
\beq
\scr{L}_{\rm eff} =
\bar\psi i \sla\d \psi
+ G (\bar\psi \psi)^2
+ \scr{O}(1/M^4),
\eeq
show that 
\beq
G = \frac{y^2}{2 M^2} \left[ 1 + \frac{c_1 y^2 \La^2}{16 \pi^2 M^2} 
+ \frac{y^2}{16\pi^2} \left( c_2 \ln \frac\La{M} + c_3 \right)\right],
\eeq
where $\La$ is the momentum space cutoff and
$c_{1,2,3}$ are numbers of order 1.
Note that all the non-analytic behavior of the loop amplitudes
cancels in the matching computation. 
Compute the fermion-fermion scattering
amplitude at one loop in the effective theory and show that it is
independent of $\La$ at one-loop order.
\eex

What is the value of $\La$ that is appropriate for a matching
calculation?
Renormalization theory tells us that the physics is
insensitive to the value of the cutoff, and so 
it does not matter.
To get complete cutoff sensitivity one should take the
cutoff to infinity, in which case the matching gives a relation
among renormalized couplings in the fundamental and effective
theory.

Note that the couplings in the effective theory are determined by
dimensional analysis in the scale $M$, provided that the dimensionless
couplings $y$ and $\la$ in the fundamental theory are order 1.
If the scale $M$ is much larger than the energy scale $E$ that is
being probed, then the effects of the higher-order effects in
the $1/M^2$ expansion are very small.

% ----------------------------------------------------------------
\subsection{Relevant, Irrelevant, and Marginal Operators}
The features we have seen in the example above are very general.
A general effective Lagrangian is defined by the particle (field)
content and the symmetries.
An effective Lagrangian in principle
contains an infinite number of operators,
but only a finite number of them are important at a given order
in the expansion in $1/M^2$, where $M$ is the scale of new physics.
It is therefore
useful to classify the terms in the effective Lagrangian according to
their dimension.
An operator of dimension $d$ will have coefficient
\beq
\De\scr{L}_{\rm eff} \sim \frac{1}{M^{d - 4}} \scr{O}_d.
\eeq
Inserting this term as a first-order perturbation to a given amplitude,
we obtain an expansion of the form
\beq
\scr{A} \sim \scr{A}_0 \left[ 1 + \frac{E^{d - 4}}{M^{d - 4}} + \cdots
\right],
\eeq
where $E$ is the kinematic scale of the physical process.
We see that the effects of operators with dimension $d > 4$ decrease
at low energies, and these are called `irrelevant' operators.
The effects of operators with $d = 4$ are independent of energy,
and these operators are called `marginal.'
Finally, the effects of operators with $d < 4$ increase with energy,
and these operators are called `relevant.'
Loop effects change the scaling behavior of couplings.
For weakly coupled theories, the most important change is that
dimensionless couplings run logarithmically.
The effects of quantum corrections on scaling behavior can be
much more dramatic in strongly-coupled conformal field theories.
See the lectures by Ann Nelson at this school.

For irrelevant and marginal operators, it makes sense for their
coefficient to be fixed by dimensional analysis at the scale $M$.
But if a relevant operator has a coefficient fixed by dimensional
analysis at the scale $M$, then its effects are not small below the
scale $M$.
In this case, the simple picture based on dimensional analysis
cannot be correct.

% ----------------------------------------------------------------------
\subsection{Naturally Small Parameters and Spurions}
In the toy model considered above,
there is a relevant operator that could be
added to the effective Lagrangian, namely a fermion mass term
\beq[toyfermmass]
\De\scr{L} = m \bar\psi \psi.
\eeq
However, this was forbidden by the chiral symmetry \Eq{discsymm}, and so
it is natural to omit this term.
As we now discuss, this symmetry also means that it is natural
for the mass to be nonzero but very small, \ie $m \ll M$.

Suppose that we modify the theory by adding the fermion mass term
\Eq{toyfermmass}.
Now the chiral symmetry \Eq{discsymm} is broken, but the effects
of this breaking come only from the parameter $m$.
Therefore, if we compute loop corrections to the fermion mass,
we find that they are proportional to the scale $m$ itself.
For example, computing the effective fermion mass by matching at
one loop gives
\beq
m_{\rm eff} = m + \frac{y^2}{16\pi^2} c m \ln\frac{\La}{M},
\eeq
where $c \sim 1$.
The size of the loop correction is controlled by $m$ rather than $M$
because an insertion of $m$ is required to break the symmetry.

This can be formalized in the following way.
We can say that the mass parameter $m$ transforms under the discrete
symmetry \Eq{discsymm} as
\beq[mtrans]
m \mapsto -m.
\eeq
What this means is that if we view $m$ as a parameter, all expressions
must depend on $m$ in such a way that the chiral symmetry including
the transformation \Eq{mtrans} is a good symmetry.
We can think of $m$ as a field, and the numerical value of $m$ as a
vacuum expectation value for the field.
We say that $m$ is a `spurion field.'
This spurion analysis immediately tells us that quantities that are
even under the chiral symmetry will depend only on even powers of $m$,
while quantities that are odd will depend on odd powers of $m$.
This kind of spurion analysis will be very useful when we consider
SUSY breaking.

% ----------------------------------------------------------------------
\subsection{Fine Tuning in a Toy Model}
To illustrate the problem with relevant operators that are not forbidden
by any symmetry, let us
consider another example of a light real scalar field $\phi$
coupled to a heavy Dirac fermion $\Psi$:
\beq
\scr{L} = \sfrac 12 (\d \phi)^2 - \sfrac 12 m^2 \phi^2
- \frac{\la}{4!} \phi^4
+ \bar\Psi i \sla\d \Psi - M \bar\Psi \Psi
+ y \phi \bar\Psi \Psi.
\eeq
Note that the mass operator $\phi^2$ has $d = 2$, and is therefore relevant.
The mass of the scalar cannot be forbidden by any obvious
symmetry, but we simply assume that $m^2$ is chosen to make the scalar
lighter than the fermion.
We can describe processes with energies $E \ll M$ using an
effective theory where the heavy fermion has been integrated out.
Carrying out the one-loop matching calculation, we find that the scalar
mass in the effective theory has the form
\beq
m^2_{\rm eff} = m^2 + \frac{y^2}{16\pi^2} \left[ c_1 \La^2
+ c_2 m^2 \ln \frac{\La}{\mu}
+ c_3 M^2 + \scr{O}(M^4 / \La^2) \right].
\eeq
where $\La$ is the cutoff used.
If we use dimensional regularization in $4 - \ep$ dimensions
and minimal subtraction, then we obtain
\beq
m^2_{\rm eff} = m^2 + \frac{y^2}{16\pi^2} \left[ \frac{c_2}{\ep} m^2
+ c_3 M^2 + \scr{O}(\ep) \right].
\eeq
In either case, we can write this in terms of the renormalized mass
$m^2(\mu = M)$:
\beq[tunex]
m^2_{\rm eff}(\mu = M) = m^2(\mu = M)
+ \frac{c_3 y^2}{16\pi^2} M^2.
\eeq
In this expression, the cutoff dependence has disappeared, but the
dependence on the (renormalized) mass $M$ remains.
This shows that if we want to make the scalar light compared to the 
scale $M$, we must tune
the renormalized couplings in the fundamental theory so that there
is a cancellation between the terms on the \rhs of \Eq{tunex}.
The accuracy of this fine tuning is of order
$y^2 m^2 / (16\pi^2 M^2)$.
There is no obvious symmetry or principle
that makes the scalar naturally light in this model.

% -----------------------------------------------------------------------
\subsection{Fine Tuning Versus Quadratic Divergences}
We see that we must fine-tune parameters whenever the low-energy
effective Lagrangian contains a relevant operator that cannot be
forbidden by symmetries.
The naturalness problem for scalar mass parameters is often said
to be a consequence of the fact that scalar
mass parameters are quadratically divergent in the UV.
We have emphasized above that the fine-tuning can be formulated
in terms of renormalized quantities, and has nothing to do with
the regulator used.
(In particular, we have seen in the example above that
fine tuning can be present in dimensional regularization,
where there are no quadratic divergences.)
The naturalness problem
is simply the fact that relevant operators that are not forbidden
by symmetries are generally sensitive to heavy physical
thresholds in the theory.

Although fine-tuning can be formulated without reference to UV
divergences, there is a close connection that is worth commenting on.
We can view a regulator for UV divergences as a UV modification of the
theory that makes it finite.
The dependence on the cutoff $\La$ can therefore be viewed as dependence
on a new heavy threshold.

% -----------------------------------------------------------------------
\subsection{Fine Tuning Versus Small Parameters}
Not all small parameters are finely tuned.
We have seen in the first toy model above that a small fermion mass is not
fine tuned, because there is an additional symmetry that results when the
mass goes to zero.
This automatically ensures that the radiative corrections to the fermion mass
are small.
In this case, we say that the small parameter is `protected by a symmetry.'

There is another general mechanism by which a parameter in an effective
Lagrangian can be naturally small, and that is if two sectors of the
theory completely decouple as the parameter is taken to zero.
As an example, consider a theory of two real scalars with Lagrangian
\beq
\!\!\!\!\!\!\!\!
\scr{L} = \sfrac 12 (\d\phi_1)^2 - \sfrac 12 m_1^2 \phi_1^2
+ \sfrac 12 (\d\phi_2)^2 - \sfrac 12 m_2^2 \phi_1^2
- \frac{\la_1}{4!} \phi_1^4
- \frac{\la_2}{4!} \phi_2^4
- \frac{\ka}{4} \phi_1^2 \phi_2^2.
\eeq
We have forbidden terms that with odd powers of the scalar fields with
discrete symmetries
\beq
\phi_1 \mapsto -\phi_1,
\qquad
\phi_2 \mapsto \phi_2,
\eeq
and
\beq
\phi_1 \mapsto \phi_1,
\qquad
\phi_2 \mapsto -\phi_2.
\eeq
This symmetry also forbids mixing terms of the form $\phi_1 \phi_2$.
If we take $\ka \to 0$, the theory becomes the sum of two `superselection
sectors,' \ie two theories
that are not coupled to each other.
It is therefore natural to take $\ka \ll \la_1, \la_2$.
It is also easy to see that any radiative correction to the
coupling $\ka$ is proportional to $\ka$ itself.
In this case, we say that the parameter $\ka$ is small because of an
`approximate superselection rule.'

Approximate superselection rules explain why it is natural for the
electromagnetic coupling to be weaker than the strong coupling.
If we take the electromagnetic coupling to zero, the theory splits
into superselection sectors, consisting of QCD and a free photon.
Approximate superselection rules also explain why it is natural
for gravity to be much weaker
than the standard model gauge interactions.%
\footnote{%
Gravity couples to all forms of matter with universal strength,
and therefore sets the ultimate limit on how decoupled two approximate
superselection sectors can be.}

% -----------------------------------------------------------------------
\subsection{To Tune or Not to Tune?}
Is fine tuning really a problem?
If we want to explain the effective couplings of the standard model
in terms of a more fundamental underlying theory, then it is at
least disturbing that the underlying couplings must be 
adjusted to fantastic accuracy in order to reproduce even the
qualitative features of the low-energy theory.
A fine tuned theory is like finding a pencil balancing on its tip:
it is possible that it arises by accident, but one
suspects that there is a stabilizing force.

Recently, the possibility that the standard model may be fine tuned
has received renewed attention, motivated in part
by the fact that there is
another grave naturalness problem: the cosmological constant
problem.
In general relativity, there is an additional relevant
operator that must be added to the Lagrangian, namely the unit operator:
\beq[cc]
\De\scr{L}_{\rm cc} = -\La^4.
\eeq
This can be thought of as a constant vacuum energy,
which is not observable in the absence of gravity.
In the presence of gravity, \Eq{cc} is covariantized to
\beq
\De \scr{L}_{\rm cc} = 
-\sqrt{\det(g_{\mu\nu})}\, \La^4,
\eeq
where $g_{\mu\nu}$ is the metric field.
This coupling means that 
vacuum energy couples to gravity.
The term \Eq{cc} gives rise to the infamous cosmological constant term in
Einstein's equations, which gives rise to a nonzero
spacetime curvature at a length scale
\beq
L \sim \frac{\MP}{\La^2}.
\eeq
In order to explain the present universe, $L$ must be at least of order
the size of the present Hubble horizon, $L_{\rm Hubble} \sim 10^{32}~{\rm cm}
\sim 10^{-42} \GeV^{-1}$.
This requires $\La \lsim 10^{-3}\eV$.

This is an enormous
problem, because loops of particles with mass $M$ give rise
to a correction to the vacuum energy of order
\beq
\De\scr{L} \sim \frac{1}{16\pi^2} M^4.
\eeq
For $M \sim M_Z \sim 100$~GeV, this is too large by 54 orders of
magnitude!
No one has ever found a symmetry that can cancel this contribution to
the required accuracy.

This problem has prompted some physicists to consider the possibility
that there could be a kind of anthropic selection
process at work in nature. 
The idea is that there are in some sense many universes with
different values of the effective couplings, and we live in one
of the few that are compatible with our existence.
If the cosmological constant were much larger than $\La \sim 10^{-3}\eV$,
then structure could not form in the universe \cite{Weinberg}.
The fact that cosmological observations favor a value of the
cosmological constant in this range has given added impetus to this
line of thinking.
Also, string theory appears to have a large number of possible
ground states, as required for anthropic considerations to operate.
For recent discussions, see \eg\ \Refs{anthropic}.

There is another possibility to save naturalness as we know it,
advocated in \Ref{Sundrumcc}.
The point is that although we have tested the standard model to
energies of order 100~GeV, we have only tested gravity to a much
lower energy scale, or much longer distances.
The shortest distances probed in present-day gravitational force
experiments are presently of order $0.1~$mm, corresponding to an
energy scale of $10^{-3}\eV$.
If we assume that new gravitational physics comes in at the $0.1$~mm
scale, the small value of the cosmological constant may be natural.
Note that this approach also predicts that the cosmological constant
should be nonzero and close to its experimental value.
The difficulty with this approach is that it is not known how to modify
Einstein gravity in a consistent way to cut off the contributions
to the cosmological constant.
However, given our ignorance of UV completions of gravity, we should
perhaps keep an open mind.
See \Ref{KS} for a recent idea along these lines.

These are interesting ideas, and worth pursuing.
But in these lectures I will assume that naturalness is a good guide
to non-gravitational physics at least.

% ------------------------------------------------------------------
\section{Model-building Boot Camp}
% ------------------------------------------------------------------
We are now ready to start building effective field theory models.
If we believe in the naturalness principle articulated in the previous
section, then the models should be defined
by specifying the particle content and the symmetries of the theory.
Then we should write down all possible couplings consistent with
the symmetries.

\subsection{The Standard Model}
Let us apply these ideas to the standard model.
The standard model is defined to be a theory with gauge group
\beq
SU(3)_C \times SU(2)_W \times U(1)_Y.
\eeq
The fermions of the standard model can be written in terms of
2-component Weyl spinor fields as%
\footnote{We use the spinor conventions of Wess and Bagger \cite{WB},
which have become conventional in the SUSY literature.
See \eg \Ref{martin} for a pedagogical introduction.}
\beq\bal
Q^i &\sim ({\bf 3}, {\bf 2})_{+\frac 16},
\\
(u^\c)^i &\sim (\bar{\bf 3}, {\bf 1})_{-\frac 23},
\\
(d^\c)^i &\sim (\bar{\bf 3}, {\bf 1})_{+\frac 13},
\\
L^i &\sim ({\bf 1}, {\bf 2})_{-\frac 12},
\\
(e^\c)^i &\sim ({\bf 1}, {\bf 1})_{+1},
\eal\eeq
where $i = 1, 2, 3$ is a generation index.
In addition, the model contains a single scalar multiplet
\beq
H \sim (1, {\bf 2})_{+\frac 12}.
\eeq

According to the ideas above, we must now write the most general
interactions allowed by the symmetries.
The most important interactions are the marginal and relevant ones.
The marginal interactions include kinetic terms for the Higgs
field, the fermion fields, and the gauge fields:
\beq
\scr{L}_{\rm kinetic} = (D^\mu H)^\dagger D_\mu H
+ Q^\dagger_i i \tilde{\si}^\mu D_\mu Q_i + \cdots
- \sfrac 14 B^{\mu\nu} B_{\mu\nu} + \cdots
\eeq
Note that these include the gauge self interactions.
Also marginal is the quartic interaction for the Higgs
\beq
\De\scr{L}_{\rm quartic} = -\frac{\la}{4} (H^\dagger H)^2
\eeq
and Yukawa interactions:
\beq
\De\scr{L}_{\rm Yukawa} = (y_u)_{ij} Q^i H (u^\c)^j
+ (y_d)_{ij} Q^i H^\dagger (d^\c)^j
+ (y_e)_{ij} L^i H^\dagger (e^\c)^j.
\eeq
Note that the Yukawa interactions are the only interactions that
break a $SU(3)^5$ global symmetry that would otherwise act on the
generation indices of the fermion fields.
This means that the Yukawa interactions can be naturally small
without any fine tuning.
This is reassuring, since it means that the 
small electron Yukawa coupling $y_e \sim 10^{-5}$
is perfectly natural.

Finally, the marginal interactions include `vacuum angle' terms for
each of the gauge groups:
\beq
\!\!\!\!\!\!
\scr{L}_{\rm vacuum\,angle}
= \frac{g_1^2 \Th_1}{16\pi^2} \tilde{B}^{\mu\nu} B_{\mu\nu}
+ \frac{g_2^2 \Th_2}{8\pi^2} \tr( \tilde{W}^{\mu\nu} W_{\mu\nu} )
+ \frac{g_3^2 \Th_3}{8\pi^2} \tr( \tilde{G}^{\mu\nu} G_{\mu\nu} ),
\eeq
where $\tilde{B}^{\mu\nu} = \sfrac 14 \ep^{\mu\nu\rho\si} B_{\rho\si}$,
{\it etc\/}.
These terms break $CP$, and are therefore very interesting.
These terms are total derivatives, \eg
\beq
\tilde{B}^{\mu\nu} B_{\mu\nu} = \d^\mu K_\mu,
\qquad
K^\mu = \sfrac 12 \ep^{\mu\nu\rho\si} A_\nu F_{\rho\si}.
\eeq
This is enough to ensure that they do not give physical effects
to all orders in perturbation theory.
They can give non-perturbative effects with parametric dependence
$\sim e^{1/g^2}$, but these are completely negligible for the
$SU(2)_W \times U(1)_Y$ terms, since these gauge couplings are
never strong.
The strong vacuum angle gives rise to $CP$-violating
non-perturbative effects in QCD, most importantly the electric
dipole moment of the neutron.
Experimental bounds
on the neutron electric dipole moment require
$\Th_{3} \lsim 10^{10}$.
Explaining this small number is the `strong $CP$ problem.'
There are a number of proposals to solve the strong $CP$ problem.
For example, there may be a spontaneously broken Peccei-Quinn
symmetry \cite{PQ} leading to an axion \cite{axion},
or there may be special flavor structure at high scales
that ensures that the determinant of the quark masses is real
\cite{NelsonBarr}.

There is one relevant interaction that is allowed, namely a mass
term for the Higgs field:
\beq
\scr{L}_{\rm relevant} = -m_H^2 H^\dagger H.
\eeq
Note that mass terms for the fermions such as $L e^\c$
are not gauge singlets, and therefore forbidden by gauge symmetry.
The Higgs mass parameter cannot be forbidden by any obvious symmetry,
and therefore must be fine tuned in order to be light compared to
heavy thresholds such as the GUT scale.
For example, in GUT models there are massive gauge bosons with masses
of order $M_{\rm GUT}$ that couple to the Higgs with strength $g$,
where $g$ is the unified gauge coupling.
These will contribute to the effective Higgs mass below the GUT scale
\beq
\De m_H^2 \sim \frac{g^2 M_{\rm GUT}^2}{16 \pi^2}
\sim 10^{30} \GeV^2
\eeq
for $M_{\rm GUT} \sim 10^{16}\GeV$.
In order to get a Higgs mass of order 100~GeV we must fine tune to one
part in $10^{26}$!

We can turn this around and ask what is the largest mass threshold
that is naturally compatible with the existence of a light Higgs
boson.
The top quark couples to the Higgs with coupling strength $y_t \sim 1$,
and top quark loops give a quadratically divergent contribution to the
Higgs mass.
Assuming that this is cut off by a new threshold at the scale $M$,
we find a contribution to the Higgs mass of order
\beq
\De m_H^2 \sim \frac{y_t^2 M^2}{16\pi^2},
\eeq
which is naturally small for $M \lsim 1\TeV$.
We get a similar estimate for $M$ from loops involving
$SU(2) \times U(1)$ gauge bosons.
So the standard model is natural as an effective field theory only
if there is new physics at or below a TeV.
This is the principal motivation for the
Large Hadron Collider (LHC) at CERN, which will start operation
in 2007-2008 with a center of mass energy of $14\TeV$.
It is expected that the LHC will discover the mechanism of
electroweak symmetry breaking and the new physics that makes
it natural.

% ------------------------------------------------------------------
\subsection{The GIM Mechanism}
One very important feature of the standard model is that it violates
flavor in just the right way.
The quark mass matrices are proportional to the up-type and down-type
Yukawa couplings.
Diagonalizing the quark mass matrices requires that we perform independent
unitary transformations on the two components of the quark doublet $Q_i$.
This gives rise to the CKM mixing matrix,
which appears in the interactions of the mass
eigenstate quarks with the $W^\pm$ (`charged currents').
Crucially, the interactions with the photon and the $Z$ (`neutral
currents') are automatically diagonal in the mass basis.
This naturally explains the phenomenology of flavor-changing
decays observed in nature, including the `GIM suppression' of flavor
changing neutral current processes such as $K^0$--$\bar{K}^0$ mixing.

For our purposes, what is important is
that this comes about because the quark
Yukawa couplings are the only source of flavor violation in the standard
model.
If there were other couplings that violated quark flavor,
these would not naturally be diagonal in the same basis that diagonalized
the quark masses, and would in general lead to additional
flavor violation.
A simple example of this is a general model with 2 Higgs doublets, in
which there are twice as many Yukawa coupling matrices.

% ------------------------------------------------------------------
\subsection{Accidental Symmetries}
It is noteworthy that the standard model was completely defined by
its particle content gauge symmetries.
In particular, we did not have to impose any additional symmetries
to suppress unwanted interactions.
If we look back at the terms we wrote down, we see that all of the
relevant and marginal interactions are actually invariant under
some additional global symmetries.
One of these is baryon number, a $U(1)$ symmetry with charges
\beq
\!\!\!\!\!\!
B(Q) = \sfrac 13,
\quad
B(u^\c) = B(d^\c) = -\sfrac 13,
\quad
B(L) = B(e^\c) = B(H) = 0.
\eeq
Another symmetry is lepton number, another $U(1)$ symmetry with charges
\beq
\!\!\!\!\!\!
L(Q) = L(u^\c) = L(d^\c) = 0,
\quad
L(L) = +1,
\quad
L(e^\c) = -1,
\quad
L(H) = 0.
\eeq
These symmetries can be broken by higher-dimension
operators.
For example, the lowest-dimension operators that
violate baryon number are dimension 6:
\beq
\De\scr{L} \sim \frac 1{M^2} Q Q Q L
+ \frac{1}{M^2} u^\c u^\c d^\c e^\c,
\eeq
where the color indices are contracted using the $SU(3)_C$
invariant antisymmetric tensor.
Consistency with the experimental limit on the proton lifetime
of $10^{33}$~yr gives a bound
$M \gsim 10^{22}\GeV$.
Although this is larger than the Planck mass, these couplings
also violate flavor symmetries, and it seems reasonable
that whatever explains the small values of the light Yukawa
couplings can suppresses these operators.

A very appealing consequence of this is that if
the standard model is valid up to a high scale $M$, then
the proton is automatically long-lived,
without having to assume that baryon
number is an exact or approximate symmetry of the fundamental theory.
Baryon number emerges as an `accidental symmetry' in the sense
that the other symmetries of the model (in this case gauge symmetries)
do not allow any relevant or marginal interactions that violate
the symmetry.

\subsection{Neutrino Masses}
Lepton number can be violated by the dimension 5 operator
\beq[nuop]
\De\scr{L} \sim \frac{1}{M} (LH)(LH).
\eeq
When the Higgs gets a VEV, these gives rise to Majorana
masses for the neutrinos of order
\beq
m_\nu \sim \frac{v^2}{M}.
\eeq
In order to get neutrino masses in the interesting range
$m_\nu \sim 10^{-2}\eV$ for solar and atmospheric neutrino mixing,
we require $M \sim 10^{15}\GeV$,
remarkably close to the GUT scale.
The interaction \Eq{nuop} also has a nontrivial flavor structure,
so the actual scale of new physics depends on the nature of flavor
violation in the fundamental theory, like
the baryon number violating interactions considered above.

The experimental discovery of neutrino masses has been heralded as
the discovery of physics beyond the standard model, but it can also
be viewed as a triumph of the standard model.
The standard model \emph{predicts} that neutrino masses (if present)
are naturally small, since they can only arise from an irrelevant
operator.
We can view the discovery of neutrino masses as evidence
for the existence of a new scale in physics.
This is analogous to the discovery of weak $\be$ decay, which can
be described by an effective 4-fermion interaction with coupling
strength $G_{\rm F} \sim 1 / (100\GeV)^2$.
(Therefore, Fermi was
doing effective quantum field theory in the 1930's!)

% ------------------------------------------------------------------
\subsection{Extending the Standard Model}
The steps in constructing an extension of the standard model are
the same ones we followed in constructing the standard model above.
The model should be defined by its particle content and symmetries.
We then write down all couplings allowed by these principles.
The goal is to find an extension of the standard model that cures the
naturalness problem, but preserves the successes of the standard model
described above.

%The ideal model would cure the naturalness problem of the standard
%model, while preserving its successes.
%The ideal model therefore would agree with all experiments and 
%naturally explain the absence of flavor-changing neutral currents.
%The feature that the standard model naturally predicts small
%baryon and lepton number violation without imposing additional
%symmetries is also a nice feature, but one that we can perhaps
%live without in an otherwise successful theory.

% ------------------------------------------------------------------
\section{The Minimal Supersymmetric Standard Model}
% ------------------------------------------------------------------
We now apply the ideas of the previous section to constructing a
supersymmetric extension of the standard model.
The motivation for this is that supersymmetry can naturally explain
why a scalar is light.
This because unbroken SUSY fixes scalar and fermion masses to be the same.
Since fermion masses can be protected by chiral symmetries, the
same chiral symmetries will also protect the masses of the scalar
superpartners.

% ------------------------------------------------------------------
\subsection{Superfields and Couplings}
To construct a supersymmetric extension of the standard model,
we simply embed all fermions of the standard model into
chiral superfields, and all gauge fields into vector superfields.
The chiral superfields are therefore
\beq\bal
Q^i &\sim ({\bf 3}, {\bf 2})_{+\frac 16},
\\
(U^\c)^i &\sim (\bar{\bf 3}, {\bf 1})_{-\frac 23},
\\
(D^\c)^i &\sim (\bar{\bf 3}, {\bf 1})_{+\frac 13},
\\
L^i &\sim ({\bf 1}, {\bf 2})_{-\frac 12},
\\
(E^\c)^i &\sim ({\bf 1}, {\bf 1})_{+1},
\eal\eeq
where $i = 1, 2, 3$ is a generation index.
The Higgs scalar fields are also in chiral superfields.
If there is a single Higgs multiplet, the fermionic partners of the Higgs
scalars will give rise to gauge anomalies.
The minimual model is therefore one with two Higgs chiral superfields
with conjugate quantum numbers
\beq\bal
H_u &\sim ({\bf 1}, {\bf 2})_{+\frac 12},
\\
H_d &\sim ({\bf 1}, {\bf 2})_{-\frac 12}.
\eal\eeq

The next step is to write the most general allowed couplings
between these fields.
Let us begin with the relevant interactions.
These are
\beq[MSSMrelevant]
\De\scr{L}_{\rm relevant}
= \myint d^2\th \bigl[
\mu H_u H_d + \ka_i L^i H_u
\bigr] + \hc
\eeq
where $\mu$ and $\ka_i$ have dimensions of mass.
Right away, we have some explaining to do.
We see that SUSY allows us to write a supersymmetric mass for the
Higgs, as well as a term that mixes the Higgs with the lepton
doublets.
(Note that $L$ and $H_d$ have the same gauge quantum numbers,
so the distinction between them is only a naming convention
up to now.)
The terms $\ka_i$ can be forbidden by lepton number symmetry,
defined by
\beq[MSSMleptonnum]
\bal
L(Q) = L(U^\c) = L(D^\c) &= 0,
\quad
L(L) = +1,
\quad
L(E^\c) = -1,
\\
L(H_u) &= L(H_d) = 0.
\eal\eeq
The `$\mu$ term' can be forbidden by a $U(1)$ `Peccei-Quinn'%
\footnote{A similar symmetry plays a role in the solution of the
strong $CP$ problem by axions, as first discussed by Peccei and Quinn.}
symmetry with charges
\beq[MSSMPQ]
\bal
P(H_u) &= P(H_d) = +1,
\\
P(Q) = P(U^\c) &= P(D^\c) = P(L) = P(E^\c) = -\sfrac 12.
\eal\eeq
There are many other symmetries that we could invent to control these
terms.
The motivation for the particular symmetries given here is that they
are not violated by Yukawa interactions (see below).
The important point is that the relevant terms in \Eq{MSSMrelevant}
can be naturally zero or small due to additional symmetries.

The marginal interactions include kinetic terms for all the gauge and
chiral superfields, which we write schematically as
\beq
\scr{L}_{\rm kinetic} \sim \myint d^4\th \left[
Q_i^\dagger e^V Q^i + \cdots \right]
+ \left( 
\myint d^2\th\, W^\al W_\al + \cdots + \hc \right).
\eeq
Note that the kinetic terms have been chosen to be diagonal in the
flavor indices.
It also includes the Yukawa couplings
\beq\bal
\scr{L}_{\rm Yukawa} = \myint d^2\th\,
\bigl[ &
(y_U)_{ij} Q^i i H_u (U^\c)^j
+ (y_D)_{ij} Q^i H_d (D^\c)^j
\\
&
+ (y_E)_{ij} L^i H_d (E^\c)^j \bigr]
+ \hc
\eal\eeq
Note that the Yukawa couplings are invariant under both the lepton
number symmetry \Eq{MSSMleptonnum} and the Peccei-Quinn symmetry
\Eq{MSSMPQ}.
There are also additional Yukawa-like interactions
\beq[RPV]\bal
\scr{L}_{\rm dangerous} = \myint d^2\th \bigl[ &
(\la_{LQD})_{ijk} L^i Q^j (D^\c)^k
+ (\la_{LLE})_{ijk} L^i L^j (E^\c)^k
\\
&+ (\la_{UDD})_{ijk} (U^\c)^i (D^\c)^j (D^\c)^k
\bigr] + \hc
\eal\eeq
Once again, we have some explaining to do.
These couplings violate lepton and baryon number symmetries,
and therefore give rise to proton decay and other processes that
are not observed unless these couplings are small.\
The couplings $\la_{LQD}$ and $\la_{LLE}$ violate lepton number,
and $\la_{LQD}$ and $\la_{UDD}$ violate baryon number, defined by
\beq\bal
B(Q) = +\sfrac 13,
\quad
B(U^\c) &= B(D^\c) = -\sfrac 13,
\\
B(L) = B(E^\c) = B(H_u) &= B(H_d) = 0.
\eal\eeq
Imposing these symmetries therefore suppresses these terms.

% ----------------------------------------------------------------------
\subsection{$R$ Parity}
Another type of possible
symmetry in SUSY theories acts differently
on different components of the same supermultiplet.
For example, we can define a $U(1)_R$ symmetry acting on chiral and
vector superfields as
\beq
\Phi(\th) \mapsto e^{iR_\Phi \al} \Phi(\th e^{-i\al}),
\qquad
V (\th) \mapsto V(\th e^{-i\al}),
\eeq
where $R_\Phi$ is the `$R$ charge' of the chiral superfield $\Phi$.
From this definition, we see that the $R$ charges of the scalar and
fermion fields in $\Phi$ are
\beq
R(\phi) = R_\Phi,
\qquad
R(\psi) = R_\Phi - 1,
\eeq
while the $R$ charge of a gaugino field is $+1$.
In order for a supersymmetric Lagrangian
\beq
\scr{L} = \myint d^4\th\, K
+ \left( \myint d^2\th\, W + \hc \right)
\eeq
to be invariant, we require $R(K) = 0$ and $R(W) = +2$.

Note that if
we define a $U(1)_R$ transformation in the MSSM
where all chiral superfields have
$R = \sfrac 13$, this is automatically preserved by all renormalizable
couplings except the $\mu$ term.

Another $R$ symmetry in the MSSM is a discrete symmetry called `$R$ parity.'
It can be defined by
\beq
\Phi(\th) \mapsto \pm\Phi(-\th),
\eeq
where the sign is $-1$ for
$Q$, $U^\c$, $D^\c$, $L$, and $E^\c$,
and $+1$ for $H_u$, $H_d$.
The idea is that the observed matter fermions have $R$ parity
$+1$, while their scalar partners have $R$ parity $-1$.
Note that the gauginos also have $R$ parity $-1$, so all
superpartners have odd $R$ parity.

$R$ parity ensures that superpartners are produced in pairs,
and that the lightest $R$ parity odd particle is absolutely stable.
$R$ parity is sufficient to forbid all of the dangerous
relevant and marginal interactions in \Eqs{MSSMrelevant} and \eq{RPV}.
(In fact, the couplings in \Eq{RPV} are often called `$R$ parity violating
operators.')
Conversely, if these operators are forbidden by another symmetry,
such as $B$ and $L$ conservation,
then $R$ parity emerges as an accidental symmetry.
Unbroken $R$ parity is often taken as part of the definition of
the MSSM, but it is worth keeping in mind that $R$ parity may
only be an accidental and/or approximate symmetry.

Note that unlike the standard model,
the MSSM requires that we impose certain exact or approximate
global symmetries \emph{in addition} to the gauge symmetries
and particle content.
In this sense, we have given up the attractive automatic
explanation of the suppression of baryon and lepton number
violation in the standard model.

% --------------------------------------------------------------------------
\section{Soft SUSY Breaking}
% --------------------------------------------------------------------------
We now begin our discussion of supersymmetry breaking.
It is obvious that the world is not exactly supersymmetric, since
SUSY predicts the existence of superpartners with the same mass and
quantum numbers as existing particles.
Once SUSY is broken the masses of the superpartners can be different
from the observed particles, and must be larger than $100\GeV$ or so
to have avoided detection in accelerator experiments performed so far.
As we have already seen above, new physics at or below TeV is required in any
solution of the naturalness problem.
In SUSY the new physics is superpartners, and therefore these must be
at or below the TeV scale, and can be discovered at LHC.
If superpartners are discovered, the most important question
in particle physics will be to understand the pattern of SUSY breaking.
It is no exaggeration to say that SUSY phenomenology is 
SUSY breaking phenomenology.

A simple way to break SUSY is to break it explicitly in the
effective Lagrangian.
If we do this, we would like to ensure that the breaking terms do not
introduce power-law sensitivity to heavy thresholds (\ie `quadratic
divergences'), which must be canceled by fine-tuning.
SUSY breaking terms with this feature are called `soft breaking terms.'
This way of breaking SUSY may be {\it ad hoc}, but it does realize the
goal of constructing a natural extension of the standard model.
Also, we will see below that if SUSY is spontaneously broken at
high scales, the effective theory below the SUSY
breaking scale is a softly broken SUSY theory.

% ---------------------------------------------------------------------------
\subsection{Coupling Constants as Superfields}
To discuss soft breaking we will use a tool that will be very useful to us
throughout these lectures.
This is the idea of coupling constants as superfields.
Note that if a superfield $\Phi$ has a nonzero value $\avg\Phi$,
it does not break SUSY as long as
\beq
Q_\al \avg{\Phi} = \bar{Q}_{\dot\al} \avg{\Phi} = 0,
\eeq
where
\beq
Q_\al = \frac{\d}{\d\th^\al} - i\si^\mu_{\al\dot\al} \bar{\th}^{\dot\al}
\d_\mu
\eeq
is the SUSY generator.
In particular, a constant nonzero value of the lowest component
of a superfield does not break SUSY.
We can therefore view the coupling constants that appear in a SUSY
theory as superfields with only their lowest components nonzero.
For example, in the Wess-Zumino model
\beq[WZex]
\scr{L} = \myint d^4\th\, Z \Phi^\dagger \Phi
+ \left[ \myint d^2\th \left(
\sfrac 12 M \Phi^2 + \sfrac 16 \la \Phi^3 \right)
+ \hc \right],
\eeq
we can view $Z = Z^\dagger$ as a real superfield and
$M$ and $\la$ as chiral superfields.

% ------------------------------------------------------------------------
\subsection{Superfield Couplings in Perturbation Theory}
This simple idea
is very useful for understanding the structure of loop corrections.
For example, in the model defined by \Eq{WZex}
the 1PI effective action at one loop contains the divergent term
\beq[ZWZ]
\De\Ga_{\rm 1PI} = \myint d^4\th\, Z \left[
1 + \frac{\hat{\la}^2}{26\pi^2} \ln \frac{\La}{\mu}
+ \hbox{\rm finite} \right],
\eeq
where
\beq
\hat\la = \frac{\la}{Z^{3/2}}
\eeq
is the physical Yukawa coupling.
By direct calculation, we find that there are no further divergences
at one loop.
In particular there are no corrections to the superpotential,
even though these are allowed by dimensional analysis.
The absence of corrections to the superpotential
holds to all orders in perturbation theory.
We now show that this can be understood very easily
if we view the couplings as superfields.

Note that a one-loop correction the superpotential in the
1PI effective action would have the form
\beq[divergentW1PI]
\De \Ga_{\rm 1PI} = \myint d^2\th\,
\frac{c \la^3}{16\pi^2} \ln\frac{\La}{\mu} + \hc
\eeq
Note that we treat $\la$ as a chiral superfield, and
therefore $\la^\dagger$ cannot appear in the 1PI superpotential.
That is, the superpotential must be a holomorphic function of the couplings
as well as the chiral superfields.
For this argument, it is important that the couplings can be treated
as superfields even in the fully regulated theory.
In this theory, this can be easily done by using a higher derivative
regulator:
\beq
\scr{L} = \myint d^4\th\, Z \Phi^\dagger
\left( 1 + \frac{\Box}{\La^2} \right) \Phi
+ \cdots,
\eeq
where the cutoff $\La$ is a real superfield.
For example, the scalar propagator is modified
\beq
\frac{i}{p^2} \to \frac{i}{p^2 - p^4 / \La^2}.
\eeq
This makes loops of $\hat{Q}$ fields UV convergent.%
\footnote{This regulator also introduces a ghost, \ie a state
with wrong-sign kinetic term at $p^2 = \La^2$.
However, this decouples when we take the limit $\La \to \infty$
and does not cause any difficulties.}
Because $\La$ is a real superfield, it cannot appear in the superpotential,
immediately ruling out divergent corrections like \Eq{divergentW1PI}.

What about finite contributions?
This holomorphy of the superpotential allows us to easily show that these
are also absent to all orders in perturbation theory.
We consider a $U(1) \times U(1)_R$ symmetry with charges given below:
\beq[NRsymm]
%\begin{table}[htbp]
\centering
\begin{tabular}{|c||c|c|}\hline
& $U(1)$ & $U(1)_R$ \\\hline \hline
$\Phi$ & $+1$ & $0$ \\\hline
$M$ & $-2$ & 2 \\\hline
$\la$ & $-3$ & 2 \\\hline
\end{tabular}
%\parbox{4in}{\caption{Symmetry charges used to prove
%non-renormalization of superpotential.}}
%\end{table}
\eeq
Note that we are treating the couplings as spurions that transform
nontrivially under these symmetries.
The most general 1PI superpotential is therefore
a function of the neutral (and dimensionless) ratio $\la \Phi^3 / M \Phi^2$:
\beq
\De \Ga_{\rm 1PI} = \myint d^2\th\, M \Phi^2 f\left( \frac{\la \Phi}{M} \right)
+ \hc
\eeq
Expanding this in powers of $\la$ we obtain
\beq
\De\Ga_{\rm 1PI} \sim \myint d^2\th \left[
\frac{M^2}{\la} \Phi 
+ M \Phi^2 + \la \Phi^3 
+ \frac{\la^2}{M} \Phi^4
+ \cdots \right] + \hc
\eeq
Only the $\Phi^2$ and $\Phi^3$ terms can be present,
since the higher order terms are singular in the limit
$\la \to 0$ or $M \to 0$.
The conclusion is that the superpotential is not
corrected in this theory.

We can use the non-renormalization of the superpotential to understand
the structure of the renormalization group (RG) equations for this theory.
Since there is only wavefunction renormalization, the physical couplings
\beq
\hat\la = \frac{\la}{Z^{3/2}},
\qquad
\hat{M} = \frac{M}{Z},
\eeq
run only because of the running of $Z$.
We therefore have (exactly)
\beq
\mu \frac{d \hat\la}{d\mu} = -\sfrac 32 \ga \la_{\rm phys},
\qquad
\mu \frac{d \hat{M}}{d\mu} = - \ga \hat{M}.
\eeq
Demanding that the one loop 1PI effective action \Eq{ZWZ}
is independent of $\mu$, we obtain the anomalous dimension
\beq
\ga = \mu \frac{d\ln Z}{d\mu}
= -\frac{|\hat\la|^2}{16\pi^2}
\eeq
which summarizes the renormalization of the theory.

% ---------------------------------------------------------------------
\subsection{Soft SUSY Breaking from Superfield Couplings}
We can include SUSY breaking terms in the Lagrangian by allowing the
superfield couplings to have nonzero higher components.
For example, the Wess-Zumino model above, we can write
\beq[SUSYbreakComp]
\bal
Z &\to 1 + (\th^2 B + \hc) + \th^2 \bar{\th}^2 C ,
\\
M &\to M + \th^2 F_M,
\\
\la &\to \la + \th^2 F_\la.
\eal\eeq
Working out the potential by integrating out the auxiliary fields,
we find
\beq
V = V_{\rm SUSY} + m^2 \phi^\dagger \phi
+ \left[ \sfrac 12 A_M \phi^2 + \sfrac 16 A_\la \phi^3 + \hc \right],
\eeq
where
\beq
V_{\rm SUSY} = \left| M \phi + \sfrac 12 \la \phi^2 \right|^2
\eeq
is the supersymmetric potential, and
\beq
\bal
m^2 &= -C + |B|^2 = -[\ln Z]_{\th^2 \bar{\th}^2},
\\
A_M &= -2 ( F_M - B M ) = -2 [\hat{M}]_{\th^2},
\\
A_\la &= -2 ( F_\la - \sfrac 32 B \la) = -2[\hat\la]_{\th^2},
\eal\eeq
where $\hat{M} = M/Z$ is the physical mass.

Now let us consider the divergence structure of this theory
including the SUSY breaking terms.
The analysis in the previous subsection
showed that in the supersymmetric case, the only
divergence is in the wavefunction renormalization, 
given to one loop by \Eq{ZWZ}.
When we turn on higher components of the superfield couplings,
this divergent contribution is given by the same expression,
but now it contains SUSY breaking from the superfield couplings.
We see that the renormalization of SUSY breaking terms that can be
written as higher components of superfield couplings is 
completely fixed by the renormalization of the couplings in the
SUSY limit.
We will explore the consequences of this in the following subsections.

Are there any additional divergences in the presence of
SUSY breaking that are not present in the SUSY limit?
These can arise from couplings that vanish identically in the
SUSY limit.
We can get such couplings by taking the total superspace integral
of a chiral quantity:
\beq
\De\Ga_{\rm 1PI} = \myint d^4 \th \left[ \al_1 \Phi
+ \sfrac 12  \al_2 \Phi^2 \right] + \hc
\eeq
The couplings $\al_1$ and $\al_2$ are renormalizable by power counting,
but are total derivatives if $\al_1$ and $\al_2$ have only the lowest
component nonvanishing
(which is why we did not include them in the original Lagrangian).
However, if $\al_1$ and $\al_2$ depend on superfield couplings with
higher components, they can have nontrivial effects.
For example,
\beq[singletcounter]
\myint d^4\th\, \al_1 \Phi 
= [\al_1]_{\th^2 \bar{\th}^2} \phi + \cdots.
\eeq
As above, we can understand the possible counterterms by considering the 
$U(1)$ and $U(1)_R$ symmetries defined in \Eq{NRsymm}.
Under these, the couplings $\al_1$ and $\al_2$ transform as
\beq
%\begin{table}[htbp]
\centering
\begin{tabular}{|c||c|c|}\hline
& $U(1)$ & $U(1)_R$ \\\hline \hline
$\al_1$ & $-1$ & $0$ \\\hline
$\al_2$ & $-2$ & $0$ \\\hline
\end{tabular}
\eeq
From these symmetries, we can see that the diagram
\beq
\BoxedEPSF{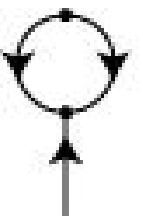}
\nonumber
\eeq
allows a logarithmically divergent contribution
\beq
\al_1 \sim \frac{\la M^\dagger}{16\pi^2} \ln\frac{\La}{\mu}.
\eeq
Note that a counterterm of the form $\al_1$ is allowed only if the
field $\Phi$ is a singlet.
The following exercise illustrates the danger of this kind of
divergence.
\bex
Consider a model of two superfields $S$ and $X$ with Lagrangian
\beq\bal
\scr{L} &= \myint d^4\th \left[
Z_S S^\dagger S + Z_X X^\dagger X \right]
\\
& \qquad
+ \myint d^2\th \left( \sfrac 12 \la S X^2 + \sfrac 12 M X^2 \right)
+ \hc
\eal\eeq
Show that in the SUSY limit $\avg{X} = 0$ while $\avg{S}$ is undetermined.
The theory therefore has a space of vacua parameterized by $S$.
Now break SUSY by turning on soft masses $m_S^2, m_X^2 \ll |M|^2$.
Show that the loop corrections destabilize the vacuum at $\avg{S} = 0$
and force
\beq
\avg{S} \sim -\frac{M}{\la}.
\eeq
\eex
Similar reasoning shows that
\beq
\al_2 \sim \frac{(\la M^\dagger)^2}{16\pi^2 M^\dagger M}
= \hbox{\rm finite}
\eeq
because there cannot be a UV divergence that is singular in the
limit $M \to 0$.

Another kind of term that can appear in the effective Lagrangian
once SUSY is broken
involves higher SUSY derivatives of the superfield
couplings, such as
\beq
D^2 \la = -4 F_\la,
\qquad
\bar{D}^2 D^2 Z = 16 C.
\eeq
However, it is not hard to check that all such terms
have positive mass dimension, and therefore cannot
be UV divergent by simple power counting.
\bex
Carry out an operator analysis to show that there are no additional
divergent counterterms involving SUSY derivatives of the couplings
in the Wess-Zumino model above.
Note that $\bar{D}_{\dot\al}$ vanishes on chiral superfields,
and that $\bar{D}^2 X$ is chiral for any superfield $X$.
\eex

We conclude that in the model \Eq{WZex}
all SUSY breaking terms that can be parameterized by a
nonzero higher component of coupling constant superfields are
soft, in the sense that they do not lead to any quadratic divergences.
Our argument has been rather abstract, and it is worth pointing out
how the the absence of quadratic divergences comes about in explicit
calculations.
In component calculations, the quadratic
divergences cancel between graphs involving loops of fermions
and bosons.
For example, the contributions to the counterterm for the scalar
mass term in the model defined by \Eq{WZex} are given in Fig.~1.
We obtain (after analytic continuation to Euclidean momenta)
\beq[loopex]
\bal
\De m^2_{\rm boson\,loop} &= +|\la|^2 \myint \frac{d^4 k_E}{(2\pi)^4}\,
\frac{1}{k_E^2 + (|M|^2 + m^2)},
\\
\De m^2_{\rm fermion\,loop} &= -\sfrac 12 |\la|^2 
\myint \frac{d^4 k_E}{(2\pi)^4}\,
\frac{\tr {\bf 1}}{k^2_E + |M|^2},
\eal\eeq
where $\tr{\bf 1} = 2$ is the trace over the Weyl fermion
indices.
We see that the quadratically divergent part cancels.
\bex
Check the signs and combinatoric factors in \Eq{loopex}.
\eex

The superfield analysis can be easily extended to include gauge fields.
To fix our superfield conventions, we write a $U(1)$ gauge superfield
as
\beq
V = V^\dagger = \cdots - 2 \th \si^\mu \bar{\th} A_\mu
+ \cdots + \th^2 \bar{\th}^2 D
\eeq
and write the field strength as
\beq
W_\al = -\sfrac 14 \bar{D}^2 D_\al V.
\eeq
We then have
\beq
\myint d^2\th\, W^\al W_\al = - 2 F^{\mu\nu} F_{\mu\nu} + \cdots + 4 D^2
\eeq
and
\beq
\myint d^4\th\, \Phi^\dagger e^V \Phi = (D^\mu \phi)^\dagger D_\mu \phi
+ \phi^\dagger D \phi + \cdots,
\eeq
where $\phi$ is the scalar component of the chiral superfield $\Phi$,
and
\beq
D_\mu = \d_\mu - i A_\mu.
\eeq
We therefore write the action of scalar QED as
\beq\bal
\scr{L} = \myint & \,d^4\th\, Z \bigl[ \Phi^\dagger e^V \Phi
+ \tilde{\Phi}^\dagger e^{-V} \tilde{\Phi} \bigr]
\\
&+ \myint d^2\th\, \frac{\tau}{8} W^\al W_\al + \hc,
\eal\eeq
where
\beq[taudef]
\tau = \frac{1}{2 g^2} - \frac{i \Th}{16\pi^2}
- \th^2 \frac{m_\la}{g^2}
\eeq
is a chiral superfield that contains the gauge coupling $g$,
vacuum angle $\Th$,
and gaugino mass $m_\la$.
The gauge kinetic term is
\beq
\myint d^2\th\, \frac{\tau}{8} W^\al W_\al + \hc
= -\frac{1}{4 g^2} F^{\mu\nu} F_{\mu\nu} + \cdots,
\eeq
so the canonically normalized gauge field is
\beq
\hat{A}_\mu = g A_\mu.
\eeq
A general renormalizable SUSY theory can be written in superfields as
\beq[genrenormSUSY]
\bal
\scr{L} &= \myint d^4 \th\, Q^\dagger_a \left( Z e^{V_A T_A} \right)^a{}_b Q^b
\\
&\qquad
+ \myint d^2\th\, \frac{\tau_A}{8} \, W^\al_A
W^{\vphantom\al}_{\al A} + \hc
\\
&\qquad 
+ \myint d^2 \th\, W(Q) + \hc,
\eal\eeq
where $a, b, \ldots$ are field indices (including both gauge and
flavor indices), $A, B, \ldots$ are gauge generator indices,
and the superpotential $W$ is a cubic function of the fields.
\beq
W = \ka_a Q^a + \sfrac 12 M_{ab} Q^a Q^b
+ \sfrac 16 \la_{abc} Q^a Q^b Q^c.
\eeq
\bex
Check that the $D$ term potential in the general theory above is given by
(for $Z = 1$)
\beq
V_D = \sum_A \frac{g_A^2}{2} (Q^\dagger T_A Q)^2
\eeq
\eex
Using the same arguments as above, we can see that turning on
nonzero higher components of these fields breaks SUSY softly,
in the sense that there are no quadratic divergences in the theory.
If there are singlets, then there may be logarithmic divergences
of the type found in \Eq{singletcounter} that may destabilize the
desired vacuum.

Is this the most general soft SUSY breaking?
We can answer this question by again using higher components
of superfield couplings to break SUSY.
Any term that breaks SUSY can be written in this way.
Consider for example the term
\beq
\De\scr{L} = \myint d^4\th\, \zeta 
D^\al \Phi i \si^\mu_{\al\dot\al}
\d_\mu \bar{D}^{\dot\al} \Phi
= 2 i [\zeta]_{\th^2 \bar{\th}^2} 
\psi^\al i \si^\mu_{\al\dot\al} 
\d_\mu \bar{\psi}^{\dot\al} + \cdots
\eeq
changes the coefficient of the fermion kinetic term relative to the
scalar kinetic term, and therefore gives rise to SUSY breaking
perturbations in the physical couplings of the
canonically normalized fermion relative to the scalar.
The superfield coupling $\zeta$ has mass dimension $-2$,
and at one loop we find divergent contributions of the form
\beq
\De\Ga_{\rm 1PI} \sim \myint d^4\th\,
\frac{\zeta |\la|^2}{16\pi^2} \La^2 \Phi^\dagger \Phi
\sim \frac{|\la|^2 [\zeta]_{\th^2 \bar{\th}^2}}{16\pi^2}
\, \La^2 \phi^\dagger \phi + \cdots.
\eeq
We see that this gives rise to a quadratically divergent
contribution to the scalar mass.
In terms of diagrams, the couplings of the scalar and fermion in
\Eq{loopex} are no longer equal, and the quadratic divergences
no longer cancel.

A more subtle example is
\beq[Cterms]
\De\scr{L} = \myint d^4\th\, \sfrac 12 C \Phi^\dagger \Phi^2 
+ \hc
= \sfrac 12 [C]_{\th^2 \bar{\th}^2} \phi^\dagger \phi^2
+ \hc + \cdots.
\eeq
The superfield coupling $C$ has mass dimension $-1$,
and at one loop we find divergent contributions of the form
\beq[Cquaddiv]
\De\Ga_{\rm 1PI} \sim \myint d^4\th\, \frac{C}{16\pi^2}
\La^2 \Phi + \hc
\sim \frac{[C]_{\th^2 \bar{\th}^2}}{16\pi^2}\,
\La^2 \phi + \hc + \cdots.
\eeq
We see that this term is not soft in general.
However, the quadratic divergence \Eq{Cquaddiv}
is absent if there are no singlets in the theory,
so that the tadpole is not allowed.
For example, a term of the form
\beq
\De\scr{L} = \myint d^4 \th\, \sfrac 12 C \Phi_1^\dagger \Phi_2^2 + \hc
\eeq
is soft if there is a $U(1)$ symmetry with charges
$\scr{Q}(\Phi_1) = 2 \scr{Q}(\Phi_2)$.
Although SUSY breaking terms of the form \Eq{Cterms}
are soft, they are usually neglected.
We will see that they are not naturally generated in the more fundamental
theories of SUSY breaking that we consider.

The result is therefore that the most general soft SUSY breaking terms 
are precisely those that can be written as higher components of superfield
couplings in the Lagrangian, plus possible `$C$' terms.
of the form \Eq{Cterms}.

% ------------------------------------------------------------------------
\subsection{Renormalization Group Equations for Soft SUSY Breaking}
The fact that the divergences in this theory are controlled completely
by the divergences in the SUSY limit gives rise to nontrivial
relations between the renormalization of SUSY and SUSY breaking
couplings.
For example, the soft scalar mass is given by
\beq
m^2 = -[ \ln Z ]_{\th^2 \bar{\th}^2}.
\eeq
\bex
Check that this formula is correct,
even for the case where the lowest component
of $Z$ is nonzero.
\eex
Therefore,
\beq
\mu \frac{d m^2}{d \mu} &= -[ \ga ]_{\th^2 \bar{\th}^2}
\\
&= \frac{1}{16\pi^2} \left[ \frac{|\la|^2}{Z^3} \right]_{\th^2\bar{\th}^2}
+ \cdots
\\
&= \frac{1}{16\pi^2} \left(
|A_\la|^2
+ 3 m^2 \right) + \cdots
\eeq
where
\beq
A_{\la} = -\frac{F_\la + 3 B \la}{Z^{3/2}}.
\eeq
We see that the RG equations for the SUSY breaking
parameters are completely determined in terms of the supersymmetric
ones.

We can easily extend these results to gauge theories.
The one loop renormalization
group equation for a $U(1)$ gauge theory is
\beq[taubeta]
\mu \frac{d \tau}{d \mu} = - \frac{\tr{\scr{Q}^2}}{16\pi^2},
\eeq
where $Q$ is the $U(1)$ charge matrix of chiral superfields.
For a non-abelian gauge
theory with gauge group $SU(N)$ and $F$ fundamentals,
the one-loop beta function is
\beq[taubeta2]
\mu \frac{d \tau}{d\mu} = \frac{3 N - F}{16\pi^2},
\eeq
Note that this immediately implies the RG equation
for the gaugino mass
\beq[RGgaugino]
\mu \frac{d}{d\mu} \left( \frac{m_\la}{g^2} \right) = 0.
\eeq

There is a subtlety in the gauge coupling superfield beyond one loop.
It is not hard to see that the 
RG equation \Eq{taubeta2} for the gauge
coupling superfield has no corrections to all orders in perturbation
theory.
This follows from the fact that $\tau$ is a chiral superfield,
and therefore the beta function
for $\tau$ must be a holomorphic function of $\tau$:
\beq
\mu \frac{d \tau}{d\mu} = \be(\tau).
\eeq
The RG equation for the real part of $\tau$ must be independent of the
vacuum angle $\Th$, which is a total derivative and therefore irrelevant in
perturbation theory:
\beq
\frac{\d \be}{\d \Im\tau} = 0.
\eeq
But because $\be$ is holomorphic, this implies that
\beq
\frac{\d \be}{\d \tau} = 0,
\eeq
\ie $\be$ is a constant.
(It can be similarly shown that $\be$ is independent of
Yukawa couplings by considering $U(1)$
charges under which the Yukawa couplings are charged spurions.)
We conclude that holomorphy implies that the one-loop RG equation
\Eq{taubeta2} is in fact valid to all orders in
perturbation theory.

This does not contradict the fact that the physical
gauge coupling does run at two loops and beyond because the
physical gauge coupling differs from the gauge coupling defined by
$\Re(\tau)$ beyond one loop.
The gauge coupling defined by the lowest component of $\tau$ is often
called the holomorphic gauge coupling to distinguish it from the
physical gauge coupling.
The physical gauge coupling is the lowest component of a real
superfield defined by
\beq[Rdef]
R = \tau + \tau^\dagger + \frac{N}{8\pi^2} \ln R
- \frac{F}{8\pi^2} \ln Z
+ \scr{O}(1/R)
\eeq
where the terms of order $1/R$ and higher are scheme dependent and
\beq
R = \frac{1}{g_{\rm phys}^2} + \left( \th^2 
\frac{m_{\la,{\rm phys}}}{g_{\rm phys}^2} + \hc \right)
+ \cdots.
\eeq
Differentiating this expression, we obtain the famous expression for the
beta function first written down in \Ref{NSVZ}
\beq
\mu \frac{d}{d\mu} \left( \frac{1}{g^2_{\rm phys}} \right)
= \frac{1}{8\pi^2} \,
\frac{\displaystyle 3N - F - F \ga}
{\displaystyle 1 - \frac{N}{8\pi^2} g^2_{\rm phys}
+ \scr{O}(g^4_{\rm phys})},
\eeq
where
\beq
\ga = \mu \frac{d \ln Z}{d\mu}.
\eeq
For a complete discussion with many applications, see \Ref{AGLR}.

% ------------------------------------------------------------------------
\subsection{Soft SUSY Breaking in the MSSM}
We now apply these results to the MSSM.
We assume $R$ parity (or equivalent symmetry) so that the 
`$R$ parity violating' terms in \Eq{RPV}
and the $H_d$--$L$ mixing terms in \Eq{MSSMrelevant} are absent.
The most general soft terms are as follows.
There are gaugino masses for the $SU(3)_C \times SU(2)_W \times U(1)_Y$
gauginos:
\beq
\De\scr{L}_{\rm gaugino} = 
- M_1 \la_1 \la_1 - M_2 \la_2 \la_2 - M_3 \la_3 \la_3 + \hc
\eeq
These can be thought of as arising from the $\th^2$ component of the
gauge coupling superfield as in \Eq{taudef}.%
\footnote{Beyond one loop, the physical gaugino mass is defined by
the $\th^2$ component of the real gauge superfield \Eq{Rdef}.
See \Ref{AGLR}.}
There are scalar masses for all scalars that can be thought
of as arising from the $\th^2 \bar{\th}^2$ component of the kinetic
coefficients:
\beq\bal
\De\scr{L}_{\rm scalar}
&= - m_{Hu}^2 H_u^\dagger H_u - m_{Hd}^2 H_d^\dagger H_d
\\
&\qquad
-(m_{\tilde{Q}}^2)^i{}_j \tilde{Q}^\dagger_i \tilde{Q}^j
- (m_{\tilde{U}}^2)^i{}_j \tilde{U}^{\c\dagger}_i \tilde{U}^{\c j}
+ \cdots .
\eal\eeq
Here we are using standard notation where the scalar components of
matter fields (those with even $R$ parity)
are denoted by a tilde, while the scalar components of the
Higgs superfields are given the same name as the superfield itself.
There are also `$A$' and `$B$' terms that can be thought of as
arising from higher components of superfield couplings, or
$\th^2$ components of kinetic coefficients:
\beq
\De\scr{L}_B &= -B\mu H_u H_d + \hc
\\
\De\scr{L}_A &= -(A_U)_{ij} \tilde{Q}^i H_u \tilde{U}^{\c j}
+ \cdots + \hc
\eeq
The names of these couplings have become traditional.
Finally, there are cubic interactions arising from the `$C$' terms
of the form \Eq{Cterms}:
\beq
\De\scr{L}_\be = (C_U)_{ij} \tilde{Q}^i H_d^\dagger \tilde{U}^{\c\dagger}
+ \cdots + \hc
\eeq
These are soft because they do not involve singlet fields.
These are usually neglected, and we will see that they do not arise in
any of the models of SUSY breaking that we consider.
For an interesting possible application of these terms, see \Ref{thomasA}.

What is the status of the MSSM with general soft breaking terms?
First of all, we should note that there are enough terms allowed
to give masses to all of the unobserved superpartners.
This is obvious for the gauginos and squarks and sleptons.
For the Higgs sector this requires some work.
See \eg the lectures by H.~Haber at this school.
This must be counted as a success.

On the other hand, there are an enormous number of parameters in the
theory once SUSY is broken, about 100 even if we use our freedom to
make field redefinitions to reduce the number of independent parameters.

% --------------------------------------------------------------------
\subsection{The SUSY Flavor Problem}
To make matters worse, many of the soft SUSY breaking
parameters have nontrivial flavor structure.
This means that they will in general
give an additional source of flavor mixing that
is not diagonal in the basis where the quark masses are diagonal.
This is the SUSY flavor problem.

Because the new flavor violation must be small, the scalar
masses must be dominantly flavor-independent, \eg
\beq
(m^2_{\tilde{Q}})^i{}_j = m^2_{\tilde{Q}} \de^i{}_j
+ (\De m^2_{\tilde{Q}})^i{}_j,
\eeq
with $\De m^2 \ll m^2$.
Also, the  $A$ terms must be dominantly proportional to the corresponding
Yukawa couplings, \eg
\beq
(A_U)_{ij} = a_U (y_U)_{ij} + (\De A_U)_{ij},
\eeq
with $\De A \ll A$.

The flavor-violating parts of the scalar masses and $A$ terms are
constrained to be small by observational constraints on flavor-changing
neutral current processes.
The most constraining process is $K^0$--$\bar{K}^0$ mixing.
In the standard model, this arises from the famous box diagram
\beq
\BoxedEPSF{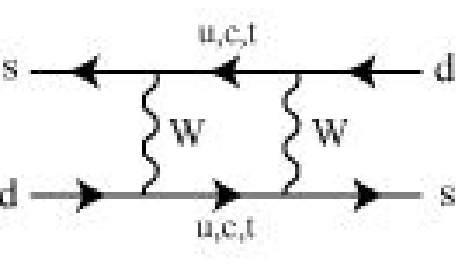}
\sim \frac{g^4}{16\pi^2} (V_{ts} V_{td})^2 \frac{m_t^2}{M_W^4}.
\eeq
In the MSSM, there are additional diagrams from squark loops.
We can treat the flavor-violating soft masses $\De m^2$ as insertions.
This gives a new contribution from box diagrams involving superpartners, \eg
\beq
\BoxedEPSF{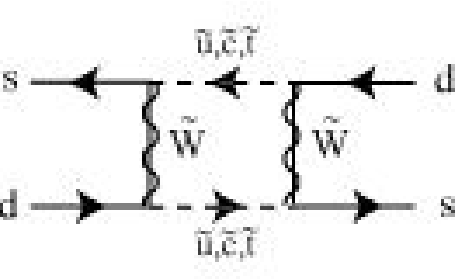}
\sim \frac{g^4}{16\pi^2} \frac{(\De m^2_{\tilde{s}\tilde{d}})^2}{m^6_{\tilde Q}}.
\eeq
Because the standard model contribution does a good job in accounting
for the observed rate, we must demand that the SUSY contribution is not larger.
This gives the bound
\beq
\frac{\De m^2_{\tilde{s}\tilde{d}}}{m^2_{\tilde{Q}}}
\lsim m_{\tilde{Q}} V_{ts} V_{td} \frac{1}{M_W}
\sim 10^{-3} \left( \frac{m_{\tilde{Q}}}{500\GeV} \right).
\eeq
We see that the squark masses have to be very nearly flavor diagonal in
order to avoid flavor-changing processes that are much larger than
what is observed.
There are many similar constraints coming from various processes.
See \Ref{Gabbiani} for a comprehensive discussion.

Because flavor symmetry is broken in the Yukawa couplings, it cannot
explain why the squark masses are nearly diagonal.
This is the SUSY flavor problem.

% ----------------------------------------------------------------------
\subsection{The $\mu$ Problem}
In order to give the Higgsino a mass
in the MSSM, we need a term of the form
\beq
\De\scr{L} = \myint d^2\th\, \mu H_u H_d + \hc
\eeq
If $\mu \gg 100\GeV$, the Higgs multiplet has a large supersymmetric
mass and electroweak symmetry cannot be broken.
If $\mu \ll 100\GeV$ the Higgsino is lighter than $M_W^2 / M_2$.
The parameter $\mu$ must therefore be of order $100\GeV$, that is,
the same size as the other SUSY breaking parameters.
Explaining why the $\mu$ term is the same size as the soft SUSY
breaking parameters is the so-called `$\mu$ problem.'
Since $\mu$ contributes to the mass term of the Higgs scalars,
we cannot explain
the weak scale without explaining the origin of the $\mu$ term.

It is not easy to understand why $\mu$ is the same size as the
soft SUSY breaking parameters discussed above because it breaks
a different set of symmetries.
For example, it does not break SUSY, but it does break
the Peccei-Quinn
symmetry defined in \Eq{MSSMPQ}.

% ----------------------------------------------------------------------
\subsection{\label{subsec:SUSYCP}The SUSY CP Problem}
Since the MSSM with SUSY breaking has many new parameters, it is not
surprising that there are new $CP$ violating phases in the
soft SUSY breaking parameters that cannot be
rotated away.

For example, there is a contribution to the strong CP phase
from the phase of the gluino mass:
\beq
\Th_{\rm QCD} = \Th_3 - \arg\det(m_u) + \arg\det(m_d) - 3\arg(M_3).
\eeq
Here $\Th_3$ is the coefficient of $\tr(F^{\mu\nu} \tilde{F}_{\mu\nu})$
in the QCD Lagrangian, and the other terms are the phases in the
masses of strongly coupled fermions.
Only the combination $\Th_{\rm QCD}$ is physically observable.
Bounds on the neutron electric dipole moment require
\beq
\Th_{\rm QCD} \lsim 10^{-10}.
\eeq
Explaining this small number is the `strong CP problem.'

The strong CP problem must be solved somehow, whether or not nature
is supersymmetric.
For example, there may be a spontaneously broken Peccei-Quinn
symmetry \cite{PQ} leading to an axion \cite{axion},
or there may be special flavor structure at high scales
that ensures that the determinant of the quark masses is real
\cite{NelsonBarr}.
These mechanisms work just as well with or without SUSY, although
in the latter case, we must independently insure that the
gluino mass does not have a phase that upsets the mechanism.
There is also an interesting proposal for solving the strong CP
problem that works only in SUSY models \cite{CPSchmaltz}.

Phases in the squark masses can also give rise to electric dipole
moments for quarks and leptons.
Demanding that the quark eletric dipole moments do not give rise
to a too-large neutron electric dipole gives the constraint
\beq
\frac{\Im \De m_{\tilde{Q}}^2}{m_{\tilde Q}^2}
\lsim 0.1 \left( \frac{m_{\tilde{Q}}}{500\GeV} \right)^2.
\eeq
Even if the SUSY breaking violates CP, this can be satisfied if
the phases are accidentally somewhat smaller than unity.
There are several constraints that are similarly strong.
See \Ref{Gabbiani} for a review.

The `SUSY CP problem' is the problem of explaining why these phases
are small.
It does not involve very large or small numbers, and is therefore
not clearly a serious problem.

% -----------------------------------------------------------------
\subsection{The SUSY Fine-Tuning Problem}
In the MSSM the Higgs potential is constrained by softly broken SUSY
to have the form
\beq[MSSMHiggsPot]
\bal
V_{\rm Higgs} &= (m_{Hu}^2 + |\mu|^2) |H_u|^2
( m_{Hd}^2 + |\mu|^2 ) |H_d|^2
- B\mu (H_u H_d + \hc)
\\
&\qquad
+ \sfrac 18 (g_1^2 + g_2^2) (|H_u|^2 - |H_d|^2)^2
+ \sfrac 12 g_2^2 | H_u^\dagger H_d |^2.
\eal\eeq
In particular, the quartic terms are completely determined, since
they arise entirely from the $D$ term potential from the
$SU(2)_W \times U(1)_Y$ gauge multiplet.

Even though the quadratic terms are free parameters, we obtain an
upper bound on the physical Higgs masses.
The basic reason is that the Higgs potential has the form
\beq
V_{\rm Higgs} \sim m_H^2 H^2 + g^2 H^4,
\eeq
so the Higgs VEV is determined to be
\beq
v \sim \avg{H} \sim \frac{m_H}{g}.
\eeq
This implies that we need $m_H \sim g v \sim M_Z$, and so the physical
Higgs mass is also of order $M_Z$.
This can be viewed as a consequence of eliminating the mass term
$m_H$ in favor of the VEV.
We might imagine that we can get larger physical higgs masses because there
are several independent quadratic terms, but famously this is not the case
for the neutral $CP$-even Higgs boson $h^0$.
Computing the physical mass of $h^0$ from
\Eq{MSSMHiggsPot} we obtain a bound%
\footnote{This assumes that the CP-odd neutral scalar $A^0$ is heavier
than $h^0$.
If this is not the case, then the bound is even stronger.}
\beq
m_{h^0} \le M_Z | \cos 2\be |.
\eeq
Here 
\beq
\tan\be = \frac{v_u}{v_d},
\eeq
where
\beq
\avg{H_u} = \pmatrix{ v_u \cr 0 \cr },
\qquad
\avg{H_d} = \pmatrix{ 0 \cr v_d \cr},
\eeq
with $v = \sqrt{v_u^2 + v_d^2} = 174 \GeV$.
A Higgs lighter than $M_Z$ was ruled out by LEP I, and the current limit
from LEP II is $m_{h^0} \ge 114\GeV$.
(It is ironic that in the standard model, the Higgs mass is in a sense
too large, while in SUSY it is too small!)

Because this bound holds independently of the Higgs quadratic terms,
we need additional contributions to the Higgs quartic couplings in order
to raise the mass of the $h^0$.
In the MSSM, these corrections can come only from loop effects.
We can compute the loop contribution to the Higgs potential
systematically by integrating out the particles heavier than the
$h^0$.
Since the heavy particles have masses and couplings that break SUSY,
the effective field theory below their masses is non-supersymmetric.
It therefore contains non-supersymmetric quartic couplings such as
\beq
\De V_{\rm Higgs} = \sfrac 14 \De\la_u (H_u^\dagger H_u)^2
+ \sfrac 14 \De\la_d (H_d^\dagger H_d)^2
+ \cdots.
\eeq
(Another way to say this is that
SUSY predicts relations among the most general allowed
quartic couplings, which are violated at loop level.)

The largest loop contribution comes from the particles with the largest
couplings to the Higgs, which are the gauge particles and the top and 
stop quarks.%
\footnote{If $\tan\be \gsim 50$, then $y_b \sim 1$ in order to explain
the observed bottom quark mass.
In this case, the sbottom can also
give a significant contribution.}
Numerically, the top contribution dominates and gives
\beq[Delau]
\De \la_u = +\frac{3 y_t^4}{4\pi^2}
\ln \frac{m_{\tilde{t}1} m_{\tilde{t}2}}{m_t^2}.
\eeq
The log can be thought of as the result of RG running between the stop
and the top mass scale.
The precise argument of the logarithm is chosen to include the finite
1-loop matching corrections.
The large size of the coefficient can be understood from the fact that
it is enchanced by a color factor.
This contributes to the physical mass of the $h^0$
\beq
\De m^2_{h^0} = \frac{3 y_t^4}{4 \pi^2} v_u^2 \sin^4\be
\ln \frac{m_{\tilde{t}_1} m_{\tilde{t}_2}}{m_t^2}.
\eeq
For numerical estimates, an important correction to the value of $y_t$
comes from the QCD corrections to the physical top quark mass:
\beq
m_t = y_t v_u \left[ 1 + \frac{g_3^2}{3 \pi^2} + \cdots \right].
\eeq
Again, the large coefficient of the 1-loop correction can be understood as
color enhancement.
For a top quark mass of $175\GeV$, we have $y_t v_u \simeq 165\GeV$.
The correction \Eq{Delau} is often written with the substitution
$y_t \to m_t / v_u$, but it is the Yukawa coupling and not the
mass that enters directly into the diagram.

The loop contribution to the Higgs quartic coupling \Eq{Delau}
grows logaritmically with the stop mass, so we can try to get a large
Higgs mass by increasing the stop mass.
However, there is a heavy price to pay for this:
the Higgs mass parameter $m_{Hu}^2$ also gets loop
contributions that grow
\emph{quadratically} with the stop mass.
In fact, this contribution is also logarithmically divergent, and
therefore has a logarithmic sensitivity to the scale $\La_{\rm SUSY}$.
These logs can be summed using the RG equations
\beq
\mu \frac{d m_{Hu}^2}{d\mu} = -\frac{3 y_t^2}{8 \pi^2}
( m^2_{\tilde{Q}3} + m^2_{\tilde{T}^\c} )
+ \cdots.
\eeq
If the logarithm is not large, we have
\beq
\De m_{Hu}^2 \simeq -\frac{3 y_t^2}{4 \pi^2}
m_{\tilde{t}}^2 \ln\frac{\La}{m_{\tilde{t}}}
\eeq
where we have assumed a common stop mass for simplicity.

It should not be surprising that the Higgs mass is quadratically sensitive
to SUSY breaking mass scales.
This is just a particular manifestation of the fine-tuning problem
for light scalars.
We argued on very general grounds that naturalness of a light Higgs
requires new physics below a TeV.
Now that we have a specific model with a specific type of new physics,
we can put in the numbers and make this more precise.
To satisfy the current experimental bounds on the Higgs mass, we require
\beq
\De m_{h^0}^2 \ge (114\GeV)^2 - M_Z^2 = (69\GeV)^2.
\eeq
We are assuming $\tan\be$, which enhances the Higgs mass.
We therefore require the stop mass to be
\beq[expbound]
m_{\tilde t} \gsim m_t \exp\left\{
\frac{2 \pi^2 \De m^2_{h^0}}{3 y_t^4 v^2} \right\}
\simeq 620\GeV.
\eeq
To quantify how much fine tuning is involved, we note that the
general scaling of \Eq{mHscaling} tells us that the natural size of the
Higgs mass parameter is of order $M_Z$.
An approximate measure of the fine tuning in the Lagrangian is therefore
\beq
\hbox{\rm tuning} &= \frac{\De m^2_{Hu}}{M_Z^2}
\simeq \frac{3 y_t^2 m_{\tilde t}^2}{4 \pi^2 M_Z^2}
\ln\frac{\La}{m_{\tilde t}}.
\eeq
This is approximately 16 for $m_{\tilde{t}} \simeq 620\GeV$ and
$\La \simeq 100\TeV$.
This means that the positive contribution to the quadratic Higgs mass
terms must cancel the large negative contribution from the loop correction
to an accuracy of at least
1 part in 16 to get acceptable electroweak symmetry
breaking (see \Eq{MSSMHiggsPot}).
The full 1-loop corrections and the largest two loop corrections
to the quartic of order $y_t^6$ and $y_t^4 g_3^2$ increase the required
stop mass above $1\TeV$ for small $A_t$, so this analysis actually
underestimates the amound of fine tuning.

This is not very much fine tuning if the parameters are at their current
experimental limits, but note that the amount of fine tuning grows
exponentially with the experimental bound on the Higgs mass
(see \Eq{expbound}).
This sensitivity also means that the precise amount of fine tuning
is sensitive to other corrections.
At present, we cannot say that there is anything clearly wrong with the
MSSM, but this may be taken as a hint that the MSSM is not a completely
natural solution to the fine-tuning problem of the standard model.

% -----------------------------------------------------------------------
\subsection{\label{subsec:NMSSM}The Next-to-Minimal
Supersymmetric Standard Model}
We can take the SUSY fine tuning problem as a motivation to
go beyond the MSSM.
A simple and well-motivated extension of the MSSM is obtained
by adding a singlet chiral superfield $S$ to the theory.

This is relevant for the SUSY fine tuning problem because
a superpotential term
\beq[NMSSMW1]
\De W = \la S H_u H_d
\eeq
gives a new contribution to the potential that is quartic in the
Higgs fields:
\beq
\De V = |F_S|^2 = |\la|^2 |H_u H_d|^2.
\eeq
This can raise the mass of the lightest neutral Higgs scalar at tree
level.
Another motivation for this model is that we can naturally obtain
nonzero values for $\avg{S}$ and $\avg{F_S}$ of order the other
SUSY breaking parameters.
This gives weak scale $\mu$ and $B\mu$ terms, solving the $\mu$ problem.

The superpotential \Eq{NMSSMW1} preserves a Peccei-Quinn symmetry
with $P(S) = -2$ (see \Eq{MSSMPQ}).
If this is broken spontaneously, it gives rise to a weak-scale axion,
which is ruled out.
This problem is easily solved by adding the superpotential
\beq
\De W_{\rm NMSSM} = \la S H_u H_d + \sfrac 16 k S^3
\eeq
to the MSSM.
This model is called the `next-to-minimal supersymmetric standard
model,' or NMSSM.

In principle, the lightest Higgs mass can be arbitrarily large by choosing
the couplings $\la$ and $k$ large, but there is a constraint if we impose
the condition that the coupling $\la$ remains perturbative up to the GUT
scale.
This is because the RG evolution
\beq
\mu \frac{d \la}{d \mu} = +\frac{\la^3}{8 \pi^2} + \cdots
\eeq
drives $\la$ larger in the UV, and $\la$ will diverge at a finite
scale if its value at the weak scale is too large.
This gives a bound of approximately $150\GeV$ on the lightest
Higgs mass in the NMSSM if we require that the theory be perturbative
up to the GUT scale $M_{\rm GUT} \simeq 2 \times 10^{16}\GeV$,
as suggested by the success of gauge coupling unification.
(Even in the standard model, requiring perturbativity of the quartic
Higgs coupling up to the GUT scale gives an upper bound of $170\GeV$ on
the Higgs mass.)
Extending the model further does not relax this bound.

There are several ways to avoid these bounds.
One possibility is that the coupling $\la$ indeed becomes strongly
coupled, so that some or all of $S$, $H_u$, or $H_d$ are composite
above the strong interaction scale.
This need not interfere with perturbative gauge coupling unification,
since the gauge couplings naturally remain weak going through a strong
threshold.
(For example, the electromagnetic coupling gets only a small
renormalization going through the QCD threshold.)
For examples of this kind of model, see \Refs{FatHiggs}.

The NMSSM phenomenology differs from the MSSM phenomenology mainly in
that scalar component of $S$ mixes with the neutral Higgs, and can
therefore change the signals of the lightest scalar.
Also, the new singlet fermion mixes with the neutralinos.

% -------------------------------------------------------------------------
\section{Spontaneous SUSY Breaking}
% -------------------------------------------------------------------------
There are good reasons to be dissatisfied with the softly broken MSSM.
It has $\scr{O}(100)$ unexplained parameters,
and correspondingly no understanding of flavor conservation,
one of the great successes of the standard model.
To address this question, it is natural to consider models in which
SUSY is broken spontaneously,
with the hope that the many SUSY breaking parameters will be
naturally explained in terms of a simpler underlying theory.
That is, we look for models with a SUSY invariant Lagrangian
in which the ground state breaks SUSY.

A famous condition for SUSY breaking comes from the fact
that the Hamiltonian can be
written in terms of the supercharges as
\beq
H = \sfrac 14 \left[
\bar{Q}_1 Q_1 + Q_1 \bar{Q}_1 + \bar{Q}_2 Q_2 + Q_2 \bar{Q}_2 \right]
\ge 0.
\eeq
This shows that the energy of any state is positive or zero,
and also that a state is SUSY invariant
\beq
Q_\al \ket\Om = 0
\eeq
if and only if its energy is exactly zero
\beq
H \ket\Om = 0.
\eeq
The vacuum energy is therefore an order parameter for SUSY breaking.
A VEV for a superfield $S$ can also act as an order parameter for
breaking SUSY.
SUSY is broken if
\beq
Q_{\al} \avg{S} = 0
\qquad
\hbox{\rm or}
\bar{Q}_{\dot\al} \avg{S} = 0.
\eeq
This occurs if higher $\th$ components of $\avg{S}$ are nonzero,
for example
\beq
\avg\Phi = \th^2 F,
\qquad
\avg{V} = \th^2\bar{\th}^2 D.
\eeq

% ------------------------------------------------------------------
\subsection{$F$-Type Breaking of SUSY}
The simplest examples of spontaneous breaking of SUSY
are O'Raifeartaigh models.
The idea behind this class of models is very simple.
Consider a theory of chiral superfields with Lagrangian of the form
\beq
\scr{L}_{\rm O'R} = \myint d^4\th\, Q^\dagger_a Q^a
+ \left( \myint d^2\th\, W(Q) + \hc \right).
\eeq
In order for there to be a vacuum with unbroken SUSY, we must have
\beq
0 = \avg{F^\dagger_a} = \Avg{\frac{\d W}{\d Q^a}}.
\eeq
If there are $N$ fields $Q^a$, this gives $N$ conditions, which
generically have a solution.
However, for special choices of $W$, there may be no solution,
and SUSY is spontaneously broken.
The order parameter is the $F$ component of a chiral superfield,
and this type of SUSY breaking is called `$F$-type SUSY breaking.'

The simplest example of this mechanism is a model with a single
superfield $S$ with superpotential
\beq
W = \ka S.
\eeq
However, this theory is trivial, as is easily seen in components.
It consists of a free chiral multiplet with a constant poential
\beq
V = |\ka|^2.
\eeq
This formally breaks SUSY, but there are no boson fermion splittings
in the model.

Suppose however that we add higher-dimension operators in the
$\int d^4\th$ terms:
\beq
\scr{L} = \myint d^4\th\, f(S^\dagger, S)
+ \left( \myint d^2 \th\, \ka S + \hc \right).
\eeq
This model is often called the `Polonyi model.'
The higher order terms in $f$ may arise from integrating out heavy
particles at scale $M$, in which case we expect
\beq[fex]
f = S^\dagger S + \frac{c}{4 M^2} (S^\dagger S)^2 + \scr{O}(S^6/M^4),
\eeq
where $c \sim 1$.
The potential terms can be computed from
\beq
\scr{L} = f_{S^\dagger S} F^\dagger F
+ ( \ka F + \hc ) + \cdots,
\eeq
where we use the abbreviation
\beq
f_{S^\dagger S} = \frac{\d^2 f}{\d S^\dagger \d S}.
\eeq
This gives
\beq
V = \frac{|\ka|^2}{f_{S^\dagger S}}.
\eeq
We see that this has a nontrivial minimum if $f_{S^\dagger S}$ has a
maximum.
For a potential of the form \Eq{fex}, we obtain $\avg{S} = 0$ for
negative $c$.
Expanding about this minimum, we find a scalar mass
\beq
m_S^2 = \frac{|c| |\ka|^2}{M^2}.
\eeq
SUSY is broken because the fermion component of $S$ remains massless.

We can write a renormalizable model of $F$-type SUSY breaking
following the original idea of O'Raifeartaigh.
One simple model contains 3 chiral superfields $S_1$, $S_2$, and
$X$, with superpotential
\beq
W = \sfrac 12 \la_1 S_1 X^2 
+ \sfrac 12 \la_2 S_2 (X^2 - v^2).
\eeq
Note that
\beq
\frac{\d W}{\d S_1} = \sfrac 12 \la_1 X^2,
\qquad
\frac{\d W}{\d S_2} = \sfrac 12 \la_2 (X^2 - v^2).
\eeq
Since these cannot both vanish, SUSY is spontaneously broken.
Since
\beq
\frac{\d W}{\d X} = (\la_1 S_1 + \la_2 S_2) X,
\eeq
the scalar potential is
\beq
V = \sfrac 14 |\la_1|^2 |X|^4
+ \sfrac 14 |\la_2|^2 |X^2 - v^2|^2
+ |\la_1 S_1 + \la_2 S_2|^2 |X|^2.
\eeq
Extremizing this potential with respect to $S_1$ and $S_2$,
it is easy to see that
\beq
\avg{\la_1 S_1 + \la_2 S_2} = 0
\quad
{\rm or}
\quad
\avg{X} = 0.
\eeq
Extremizing with respect to $X$ gives
\beq
\avg{X^2} = \frac{|\la_2|^2}{|\la_1|^2 + |\la_2|^2} v^2
\quad
{\rm or}
\quad
\avg{X} = 0.
\eeq
The value of the vacuum energy is
\beq
\avg{V} = \begin{cases}
\displaystyle \vphantom{\Biggl[}
\frac{|\la_2|^2}{4} |v|^4
& for $\avg{X} = 0$, \cr
\displaystyle \vphantom{\Biggl[}
\frac{|\la_1|^2 |\la_2|^2}{2 (|\la_1|^2 + |\la_2|^2)} |v|^4
 & for $\avg{X} \ne 0$.
\end{cases}
\eeq
It is easy to see that the vacuum energy is minimized at $\avg{X} \ne 0$
provided that $|\la_2| > |\la_1|$.
In this case $\avg{\la_1 S_1 + \la_2 S_2} = 0$, but one linear
combination of $S_1$ and $S_2$ is completely unconstrained.
\bex
Show that in an arbitrary O'Raifeartaigh model there is always
one linear combination of the superfields that is unconstrained
by minimizing the potential at tree level.
\eex

Working out the scalar and fermion masses at the minimum of the
potential, we find that all scalar and fermion masses are of
order $\la v$
except for the scalar and fermion components of the superfield
that is orthogonal to the linear combination 
$\la_1 S_1 + \la_2 S_2$.
This suggests that we can integrate out all the fields except
one chiral superfield, and write the effective Lagrangian 
below the scale $\la v$ as an effective theory of a single
chiral superfield.
This can be formally justified by noting that
\beq
\avg{F_X} &= 0,
\\
\la_1 \avg{F_{S1}} + \la_2 \avg{F_{S2}}
&= \sfrac 12 |\la_1|^2 X^{\dagger 2}
+ \sfrac 12 |\la_2|^2 (X^{\dagger 2} - v^{\dagger 2}) = 0,
\eeq
so that
\beq
X = 0,
\qquad
\la_1 S_1 + \la_2 S_2 = 0
\eeq
are superfield constraints that can be used to define the light
degrees of freedom in the effective theory.
The massless chiral multiplet can therefore be parameterized by
$S_2$ (for example), which gives the effective Lagrangian
\beq
\scr{L}_{\rm eff} = \myint d^4\th \left[
1 + \frac{|\la_2|^2}{|\la_1|^2} \right] S_2^\dagger S_2
+ \left( \myint d^2\th\, \left[ -\sfrac 12 \la_2 v^2 \right] S_2
+ \hc \right).
\eeq
This has precisely the form of the Polonyi model considered above!
In this sense, O'Raifeartaigh models reduce to Polonyi models at low
energies.

Note that in the O'Raifeartaigh model, the corrections to the effective
kinetic function $f$ come from integrating out massive fields at
the scale $\la v$, and is therefore fully calculable.
Computing the Coleman-Weinberg
potential in this model, one finds that the minimum is at
$\avg{S_2} = 0$.

% ----------------------------------------------------------------------
\subsection{$D$-Type Breaking of SUSY}
We can also break SUSY with an order parameter that is the
highest component of a gauge superfield:
\beq
\avg{V} = \th^2 \bar{\th}^2 \avg{D}.
\eeq
The simplest way works for only for a $U(1)$ gauge field, and
consists of adding a `Fayet-Iliopoulos term'
\beq
\De\scr{L}_{\rm FI} = \myint d^4\th\, \xi V,
\eeq
where $\xi$ is a coupling with
mass dimension $+2$.
This is is gauge invariant because under a gauge transformation
$\de V = \Om + \Om^\dagger$, where $\Om$ is a chiral superfield,
so under gauge transformations
\beq
\de (\De\scr{L}_{\rm FI}) = \myint d^4\th\, \xi (\Om + \Om^\dagger)
= {\rm total\ derivative}.
\eeq
However, notice that this term is not gauge invariant if the
coupling is a superfield.
This means that this term cannot be generated by a more fundamental
theory in which the couplings are superfields, such as a SUSY gauge
theory.
As we will discuss below, this also means that this term is 
not allowed by supergravity.
% (We will understand this when we study supergravity below.)
We will therefore not discuss these terms further.

It is possible to get $\avg{D} \ne 0$ if SUSY and gauge symmetry
are broken at the same time.
For example, consider a theory with superfields $Q$, $\tilde{Q}$,
$X$, and $\tilde{X}$, with $U(1)$ gauge charges
$\pm 1$ and $\pm 2$ respectively, and one singlet $S$.
The superpotential
\beq
W = \sfrac 12 \la (\tilde{X} Q^2 + X \tilde{Q}^2)
+ \la' S (Q \tilde{Q} - v^2)
\eeq
breaks SUSY and gauge symmetry at the same time, and gives rise
to a nonzero value of $\avg{D}$.
As this example illustrates, models of this type are necessarily
somewhat complicated.
An interesting open question is whether one can naturally get
$\avg{D} \gg \avg{F}$ without Fayet-Iliopoulos terms.

% --------------------------------------------------------------------
\subsection{Generalities of Tree-Level SUSY Breaking}
Let us consider a tree-level SUSY Lagrangian of the form
given in \Eq{genrenormSUSY}.
We are interested in tree-level SUSY breaking, so we take $Z = 1$
and all couplings to be SUSY preserving.
The potential can be written as
\beq
V = F^a F^\dagger_a + \sfrac 12 D_A D_A
\eeq
where
\beq
F^a = W^{\dagger a} = \frac{\d W^\dagger}{\d Q^\dagger_a},
\qquad
F^\dagger_a = \frac{\d W}{\d Q^a},
\qquad
D_A = g_A Q^\dagger_a (T_A)^a{}_b Q^b.
\eeq
(Note that we have absorbed a factor of the gauge coupling into the
definition of $D_A$.)
SUSY is spontaneously broken if $\avg{F_a}$ and/or $\avg{D_A}$ are nonzero.

We first show that if SUSY is spontaneously broken, there is a massless
fermion in the spectrum, called the `Goldstino.'
The fermion mass terms in this notation are
\beq
\scr{L}_{\rm fermion\,mass} = -\sfrac 12 \psi^a W_{ab} \psi^b
+ \sqrt{2} \la_A D_{Aa} \psi^a + \hc,
\eeq
where
\beq
D_{Aa} = \frac{\d D_A}{\d Q^a},
\eeq
\etc, and we use the notation $W_{ab} = \avg{W_{ab}}$
in the remainder of this subsection.
We can write the fermion masses in a matrix notation as
\beq
\scr{L}_{\rm fermion\,mass} = \Psi M_{\half} \Psi + \hc,
\eeq
where
\beq
\Psi = \pmatrix{\psi^a \cr \la_A \cr},
\eeq
and
\beq
M_{\half} = \pmatrix{ W_{ab} & -\sqrt{2} g_B D_{Ba} \cr
-\sqrt{2} D_{Ab} & 0 \cr}.
\eeq
We now claim that the fermion
\beq
\Psi_{\rm Goldstino} = \pmatrix{ W^{\dagger b} \cr - D_B / \sqrt{2} \cr}
\eeq
is massless.
(Note that if SUSY is unbroken, $\Psi_{\rm Goldstino}$ is trivial.)
This follows from
\beq
M_{\half} \Psi_{\rm Goldstino} = \pmatrix{ W_{ab} W^{\dagger b}
+ D_B D_{Ba} \cr 
-\sqrt{2} D_{Ab} W^{\dagger b} \cr}.
\eeq
The upper component vanishes because
\beq
0 = \frac{\d V}{\d Q^a} = W_{ab} W^{\dagger b} + D_B D_{Ba}.
\eeq
The lower component vanishes due to the gauge invariance of the superpotential:
\beq
0 = \de W = W_a \de Q^a = W_a (T_A)^a{}_b Q_b
= \frac{1}{g_A} W_a D_A^a.
\eeq
%Therefore,
%\beq
%W_a D_A^a = 0,
%\qquad
%W^{\dagger a} D_{Aa} = 0.
%\eeq

The existence of the Goldstino when SUSY is spontaneously broken
is analogous to the existence of a Goldstone boson when a global
symmetry is spontaneously broken.
Its existence can also be established on general grounds,
without referring to any Lagrangian.
For a very clear discussion, see \Ref{Witten}.

Let us continue our discussion of the masses with the scalar masses.
Combining the scalar fields into a vector
\beq
\Phi = \pmatrix{ Q^a \cr Q^\dagger_a \cr},
\eeq
we can write the scalar masses as
\beq
\scr{L}_{\rm scalar\,mass} = -\Phi^\dagger M^2_0 \Phi,
\eeq
where
\beq[mscalar]
\!\!\!\!\!
M^2_0 = \pmatrix{ W^{\dagger a c} W_{cb} + D_A^a D_{Ab} + D_{Ab}^a D_A &
W^{\dagger abc} W_c + D_A^a D_A^b \cr
W_{abc} W^{\dagger c} + D_{Aa} D_{Ab} &
W_{ac} W^{\dagger cb} + D_{Aa} D^b_A + D_{Aa}^b D_A \cr}.
\eeq
Finally, the gauge boson masses are
\beq
\scr{L}_{\rm gauge\,mass} = \sfrac 12 A^\mu M_1^2 A_\mu,
\eeq
with
\beq
(M_1^2)_{AB} = 2 D_A^a D_{Ba}.
\eeq
From these formulas, we can read off the fact that the `supertrace'
of the mass matrix vanishes:
\beq
\mathop{\rm str}(M^2) \equiv \tr(M_0^2)
- 2 \tr(M_{\half}^\dagger M^{\vphantom\dagger}_{\half})
+ 3 \tr(M_1^2) = 0.
\eeq
The supertrace is the sum of the squared masses of the particles,
counting spin multiplicities, with fermions contributing with opposite
sign as bosons.
The vanishing of the supertrace puts strong constraints on how SUSY
is broken, as we will see.

% ---------------------------------------------------------------------
\subsection{SUSY Breaking in the Observable Sector}
We have seen if SUSY makes electroweak symmetry breaking
natural, superpartner masses must be below $\sim 1\TeV$,
and to explain their non-observation they must have masses greater
than $\sim 100\GeV$.
The superpartner masses must therefore be at the weak scale, and
a good model of SUSY breaking should explain why this is so.

An obvious thing to try is to break SUSY and electroweak
symmetry at tree level by some extended Higgs sector.
That is, we imagine a renormalizable extension of the MSSM to include
extra fields and interactions that break SUSY and electroweak symmetry at tree
level.
In such a model, superpartner masses are nonzero because of direct
couplings to the fields that break SUSY.

We can see that this is difficult from the supertrace constraint
discussed in the previous subsection.
First of all, note that the mass matrix has a block-diagonal form,
with each block corresponding to states that do not mix with the
states in the other blocks.
(For example, colored particles to do not mix with color singlets.)
Unless there are heavy fermions in every block containing
observed quarks and leptons, the supertrace constraint immediately
implies that the scalar superpartners must be lighter than the
heaviest observed fermion, which is a phenomenological disaster.

It can be shown that even if we allow for the possibility
of heavy particles, 
there are always light scalar color triplets (squarks) lighter than
either $m_u$ or $m_d$, the lightest quark masses.
Let us therefore restrict attention to the color triplet part of the
mass matrix \Eq{mscalar}, which does not mix with anything else.
Because color is unbroken we have
\beq
\avg{W_a} = \avg{W^{\dagger a}} = \avg{D_{Aa}} = \avg{D_A^a} = 0
\eeq
when $a$ is a color triplet index.
The color triplet part of the scalar mass matrix is therefore
\beq
M_0^2 = \pmatrix{ M_{\half}^\dagger M^{\vphantom\dagger}_{\half}
+ \avg{D^a_{Ab} D_A} &
\De \cr 
\De^\dagger & M^{\vphantom\dagger}_{\half} M_{\half}^\dagger
+ \avg{D_{Aa}^b D_A} \cr}.
\eeq
where
\beq
\De^{ab} = \avg{W^{\dagger abc} W_c}.
\eeq
The idea is to think of $M_0^2$ as a Hamiltonian, with the lightest
mass eigenvalue equal to the ground state energy.
We can use the standard variational method from quantum mechanics
to estimate the ground state energy.
For any `state vector' $\Phi_0$, we have
\beq
\frac{\Phi_0^\dagger M_0^2 \Phi_0}{\Phi_0^\dagger \Phi_0}
\ge \hbox{\rm smallest\ eigenvalue}.
\eeq
To choose $\Phi_0$, note that the
$D$ term contribution $\avg{D^a_{Ab} D_A}$
is proportional to charges, and is therefore negative
at least one of the fields $U$, $U^\c$, $D$, and $D^\c$
in each generation (in a mass basis).
Suppose for concreteness it is $U$ that gives a negative result.
We then define
\beq
\Phi_0 = \pmatrix{\phi \cr \phi^\dagger \cr},
\eeq
where
\beq
\phi = \pmatrix{ 1 \cr 0 \cr \vdots \cr 0 \cr}
\eeq
is a unit vector in the direction of the first generation $U$ field,
with
\beq
\avg{D^a_{Ab} D_A} \phi^b = -|\la| \phi^a.
\eeq
Then
\beq
\Phi_0^\dagger M_0^2 \Phi_0 = \sfrac 12 
\phi^\dagger ( M_{\half}^\dagger M^{\vphantom\dagger}_{\half}
+ \avg{D_A g_A T_A} \phi
+ \hc ,
\eeq
where we have used the fact that $\phi \De \phi$ vanishes, since there are
no mass terms of this form $UU$ allowed.
Therefore
\beq
\frac{\Phi_0^\dagger M_0^2 \Phi_0}{\Phi_0^\dagger \Phi_0} = m_u^2 - |\la|.
\eeq
This implies that the matrix $M_0^2$ has at least one eigenvalue lighter
than the quark mass $m_u^2$.
This is clearly ruled out.
If the quark with the negative eigenvalue is down-type, we have an
eigenvalue less than $m_d^2$, which is also ruled out.

The basic problem is that $D$-type masses are proportional to charges,
and therefore have both positive and negative signs,
while the `$B$-type' masses parameterized by the off-diagonal $\De$
terms tend to split the eigenvalues, and therefore cannot raise the
lowest eigenvalue.

% -----------------------------------------------------------------------
\subsection{The Messenger Paradigm}
To make a viable model of SUSY breaking, we need either large loop
corrections, or non-renormalizable terms in the \Kahler potential.
SUSY breaking from either of these sources is suppressed,
by loop factors and/or by high mass scales.
This means that the theory must contain a sector in which SUSY is
broken at a scale much larger than 
the weak scale.
This large primordial SUSY breaking will then communicated to the
standard model fields through `messenger' interactions.

This gives us a way of thinking about the SUSY flavor
problem.
Since the interactions of the messengers with the standard model
fields determine the pattern of SUSY breaking in the visible sector,
a natural way to avoid additional flavor violation in the MSSM
is if the messenger interactions do not violate flavor symmetries,
\ie are `flavor blind.'
We will see that this paradigm gives rise to successful models of
SUSY breaking.

% -----------------------------------------------------------------------
\section{\label{sec:hidden}Hidden Sector SUSY Breaking}
%We have argued that the natural way to approach building models of
%SUSY breaking is to communicate SUSY breaking by messenger interactions.
%We can therefore classify different SUSY breaking models by the kind of
%messenger interactions.
%Since we want to avoid breaking flavor symmetries in the SUSY breaking
%sector, we look for a flavor-blind messenger interaction.

An obvious candidate for the messenger of SUSY breaking is gravity.
From a particle physics perspective, the unique low-energy
effective theory of gravity is general relativity.%
\footnote{This assumes that gravity is mediated by a spin-2 boson,
and also assumes locality and Lorentz invariance.
The cosmological constant problem has motivated
attempts to relax the assumption of locality \cite{nonlocal} or
Lorentz invariance \cite{ghostcond}.}
Its consistency requires that gravity couples to matter
through the stress-energy
tensor, which is the origin of the equivalence principle.
Because gravity couples to all forms of energy, it necessarily couples
the SUSY breaking sector with the visible sector, even if there are no
other interactions between the two sectors.
All that is required for gravity to be the messenger of SUSY breaking
is that there are no other stronger interactions between the two
sectors.
In this case, we refer to the SUSY breaking sector as the `hidden sector.'

The fact that gravity couples to the stress-energy tensor also means that
general relativity is flavor-blind.
(Different flavors have different masses, and in this sense
couple differently to gravity.
But what we want is that there be no \emph{additional} flavor
violation beyond that of the Yukawa couplings.)
The difficulty with this in practice is the fact that general relativity
is only an effective theory, and requires UV completion above the Planck
scale (if not at lower scales).
It is far from clear that the fundamental theory of gravity can have 
flavor symmetries that guarantee that the UV couplings of gravity are
flavor-blind.
In fact, there are strong hints from what little is known about the UV
theory of gravity to suggest that the fundamental theory of gravity
is unlikely to respect global symmetries such as flavor.
One of these hints comes from studies of black holes, where Hawking
radiation appears to be incapable of radiating away any global charge that
was thrown into the black hole when it was formed.
The final stages of black hole evaporation occur when the mass of the
black hole becomes of order the Planck mass, and what happens there
is not understood.
However, requiring the conservation of global quantum numbers appears to
require a large number of charged states at the Planck scale (corresponding
to all the different possible charges of the initial black hole) which seems
unlikely.
Another hint comes from string theory, the only known candidate for a
fundamental theory of gravity.
String theory does not appear to allow exact global symmetries, % \cite{},
although the full space of string theory vacua is still poorly
understood.

From a low-energy point of view, we can parameterize the most general
effects of the unknown physics at the Planck scale by higher dimension
operators suppressed by powers of the Planck scale $\MP$.
Of particular interest to us are operators that connect the fields
in the hidden sector with those in the observable sector.
We assume that SUSY is broken in the hidden sector by the $F$ component
of a field $X$, and without loss of generality we shift the field $X$
so that
\beq
\avg{F_X} \ne 0,
\qquad
\avg{X} = 0.
\eeq
We can then write the most general interactions between
$X$ and the visible sector fields:
\beq[hiddenSUSY]
\bal
\De\scr{L} &= \myint d^4\th\, \Bigl\{
\frac{(z_Q)^i{}_j}{\MP^2}\, X^\dagger X Q^\dagger Q + \cdots 
\\
&\qquad\qquad\quad
+ \frac{b}{\MP} X H_u H_d
+ \frac{b'}{\MP} X^\dagger X H_u H_d + \hc \Bigr\}
\\
&\qquad
+ \myint d^2\th \left[ \frac{s_1}{\MP} X W^\al_1 W_{1\al} + \cdots \right]
+ \hc
\\
&\qquad
+ \myint d^2\th \left[ \frac{a_{ij}}{\MP} X Q^i H_u (U^\c)^j + \cdots 
\right]
\eal\eeq
When we substitute the SUSY breaking VEV $\avg{F_X}$, we find that this
generates all the soft SUSY breaking terms of the MSSM (other than the
`$C$' terms of \Eq{Cterms}) with the size of all SUSY breaking masses of order
\beq
M_{\rm SUSY} \sim \frac{\avg{F_X}}{\MP}.
\eeq
Taking $M_{\rm SUSY} \sim \TeV$ gives
\beq
\avg{F_X} \sim \MP M_{\rm SUSY} \sim (10^{11} \GeV)^2.
\eeq
(The scale $10^{11}\GeV$ is often called the `intermediate scale.')
Note that the $\mu$ and $B\mu$ terms are generated by the terms
with coefficients $b$ and $b'$ in \Eq{hiddenSUSY}.
This gives a very simple solution to the `$\mu$ problem', as first pointed
out by \Ref{GM}.
Note these terms are only allowed if the SUSY breaking field $X$ is a singlet,
as are the terms with couplings $s_1, \ldots$ that give rise to
gaugino masses.

It is easy to see why SUSY breaking terms are suppressed in this approach.
For example, `$C$ terms' are generated by operators of the form
\beq
\De\scr{L}_{\rm eff} \sim
\myint d^4\th\, \frac{X^\dagger X}{\MP^3}\,
Q H^\dagger_d U^\c + \hc,
\eeq
which give rise to SUSY breaking trilinear couplings
of order $M_{\rm SUSY}^2 / \MP \ll M_{\rm SUSY}$.
`Hard' SUSY breaking is also small.
For example, a fermion kinetic term arises from
\beq
\De\scr{L}_{\rm eff} \sim
\myint d^4\th\, \frac{X^\dagger X}{\MP^4}\,
D^\al Q \si^{\mu}_{\al\dot\al} \d_\mu \bar{D}^{\dot\al} Q^\dagger,
\eeq
which gives $\De Z \sim M_{\rm SUSY}^2 / \MP^2 \ll 1$.
It is striking that simply writing all possible terms connecting
the hidden sector to the visible sector suppressed by powers of
a single large scale gives all required SUSY breaking terms
(including $\mu$ and $B\mu$ terms),
all of the same order.
\bex
Write the leading additional SUSY breaking allowed in the NMSSM coupled
to a hidden sector.
Does this automatically give rise to all allowed SUSY breaking of order
$M_{\rm SUSY}$, as in the MSSM?
Can we impose symmetries so that all required SUSY breaking is generated
with size $M_{\rm SUSY}$?
\eex

%Suppose that we add a term
%\beq
%\De\scr{L} \sim \myint d^4\th\, \frac{1}{\MP^2} X^\dagger (D^2 X) S
%+ \hc
%\eeq
%Show that this gives rise to a linear term in the effective
%superpotential with coefficient $M_{\rm SUSY}^2$.
%\eex

The difficulty with this approach is that the soft masses and
$A$ terms can violate flavor.
The $A$ terms arise from the terms with coefficients $a_{ij}$,
and we can imagine forbidding these by symmetries acting on
the field $X$.
However, the soft scalar masses are generated by the operators with
coefficients $z^i{}_j$, which are invariant under all symmetries.
% (In any case, we need the soft scalar masses to be present!)
Unless there are flavor symmetries at the Planck scale, there appears
to be no reason for these coefficients to be flavor-diagonal.
This is the flavor problem of hidden sector models of SUSY breaking.

One way to avoid the flavor problem is to assume that there is
a gauged flavor symmetry at the Planck scale.
The existence of gauge symmetries (as opposed to global symmetries)
is compatible with
what is known about string theory and black hole physics.
A gauge symmetry must be free of anomalies, but extra fermions can
always be added to cancel the anomalies, and once the flavor symmetries
are broken, all these extra fermions can in principle become massive.
The flavor gauge symmetry must be broken at a high scale to avoid
dangerous flavor-changing neutral currents.
The flavor symmetry may also be discrete.
For an example of this kind of model, see \Ref{barbhall}.

%M. Dine, R. Leigh and A. Kagan, Phys. Rev. D48 (1993) 4269; Y. Nir and
%N. Seiberg, Phys. Lett. B309 (1993) 337.
%1)  PREDICTIONS FROM A U(2) FLAVOR SYMMETRY IN SUPERSYMMETRIC THEORIES.
%By Riccardo Barbieri (Pisa U. & INFN, Pisa), G.R. Dvali (CERN), Lawrence J. Hall (UC, Berkeley & LBL, Berkeley),. LBL-38065, UCB-PTH-95-44, Dec 1995. 12pp.
%Published in Phys.Lett.B377:76-82,1996
%e-Print Archive: hep-ph/9512388

\bex
To illustrate some of the issues involved in models with flavor
symmetries, consider the following model.
Assume that the model at the Planck scale preserves an $SU(3)^5$
flavor symmetry that acts separately on the generation indices of
the five multiplets $Q$, $U^\c$, $D^\c$, $L$, and $E^\c$.
Find additional particle content that can make the $SU(3)^5$
flavor symmetry free of gauge anomalies, such that the additional
particles can get masses below the $SU(3)^5$ that do not violate
standard model gauge symmetry.

Suppose further that these symmetries are spontaneously broken by
the VEVs of scalar fields $Y$ that have the same quantum numbers
as the Yukawa couplings.
(For example, there are one or more `up-type' fields $(Y_U)_{ij}$, where
the $i$ is a $SU(3)_Q$ and $j$ is a $SU(3)_U$ index.)
The Yukawa couplings are of order
\beq[Yukex]
y \sim \frac{\avg{Y}}{\MP}.
\eeq
The hierarchy of VEV's for different components of $Y$ gives rise to
the hierarchy for Yukawa couplings.
Show that in this model, the off-diagonal squark masses have size
\beq
\frac{\De m^2_{\tilde Q}}{m_{\tilde Q}^2} \sim y^2,
\eeq
where $y$ is an appropriate Yukawa coupling.
Show that this is sufficient to suppress dangerous flavor-changing
neutral currents.
\eex

The model in the example above is 
complicated, and contains many particles.
The top quark Yukawa coupling is $y_t \sim 1$, which means that
the expansion in powers of $\avg{Y}/\MP$ is marginal.
(Since loop effects are suppressed by powers of $1/(16\pi^2)$,
the expansion really only breaks down for $y_t \sim 4\pi$,
so this is not fatal.)
However, all these features arise because the model is an ambitious
attempt to explain the
origin of the Yukawa couplings.
Since the Yukawa couplings presumably have \emph{some} explanation,
and no simple one is known, this may not be a drawback of this
approach.

We should note that this is not the way flavor arises in conventional
string compactifications, in which there are no flavor symmetries at the
string scale (see \Ref{GSW2} for a review).
On the other hand, string theory has not had any real success so far in
explaining the observed features of our world.
Maybe string theory is right, but we have not found the appropriate
string vacuum.
Given the richness of string theory, it might be right and we might
never know.
In this situation it is worth keeping an open mind about mechanisms
that do not fit into string theory in an obvious way.

% -----------------------------------------------------------------------
\subsection{The `Minimal SUGRA' Ansatz}
An {\it Ansatz} that has been extensively analyzed in the literature is to
assume that the couplings that give rise to scalar masses
are equal to a universal value at $\mu = \MP$:
\beq
(z_Q)^i{}_j = (z_L)^i{}_j 
= \cdots = z_0 \de^i{}_j,
\quad
z_{H_u} = z_{H_d} = z_0
\eeq
This is called `minimal SUGRA' for historical reasons.
One feature of this {\it Ansatz} is that the up-type Higgs mass runs
negative because of the large top Yukawa coupling.
This is called `radiative symmetry breaking.'
We have argued above that if there is no flavor symmetry at the Planck
scale, this {\it Ansatz} is not natural.
Nonetheless, there is an extensive literature on this, so you
should at least know what it is.
\bex
In this Ansatz, we must run the $z$ couplings down from the scale
$\MP$ down to the mass of the field $X$.
Below this scale, we match onto a theory where $X$ is replaced
by its SUSY breaking VEV.
Show that if we include only standard model fields
in the loops, this procedure is equivalent to running universal
scalar masses down from the Planck scale.
\eex

% -----------------------------------------------------------------------
\section{\label{sec:GMSB}Gauge Mediated SUSY Breaking}
% -----------------------------------------------------------------------
Following the messenger paradigm for SUSY breaking,
another natural flavor-blind messenger to consider are the standard
model gauge interactions themselves.
We have seen that tree-level SUSY breaking in the visible sector
has severe difficulties, so we look at loop effects.

% -----------------------------------------------------------------------
\subsection{Gauge Messengers}
A simple and predictive framework is to assume that SUSY breaking
is communicated to the standard model via heavy chiral supermultiplets
that are charged under the standard model gauge symmetries.
If the masses of these messenger fields are not exactly supersymmetric,
integrating them out will give rise to SUSY breaking in the visible 
sector.

A simple example that illustrates how this kind of SUSY breaking can
arise is an O'Raifeartaigh-type model with superpotential
\beq\bal
W &= \sfrac 12 \la_1 S_1 X^2 + \sfrac 12 \la_2 S_2 (X^2 - v^2)
+ \la_3 S_3 \tilde{\Phi} \Phi
\\
&\qquad + (M + \la S_2) \tilde{\Phi} \Phi.
\eal\eeq
Here $S_i$ ($i = 1, 2, 3$) and $X$ are singlet fields,
and $\Phi$ and $\tilde\Phi$ are charged under the standard model.
This model breaks SUSY, and for appropriate choices of the parameters,
the minimum occurs for
\beq
\avg{F_{S_2}} \sim \la v^2
\eeq
and $\avg{S_1}, \avg{S_3}, \avg{\Phi}, \avg{\tilde{\Phi}} = 0$.
This means that $\Phi$ and $\tilde\Phi$ effectively have a
SUSY breaking mass term, with a superfield mass parameter
\beq[messmassex]
\scr{M} = M + \la \avg{S_1} + 
\th^2 \avg{F_{S_2}}.
\eeq
As long as the fields in this sector do not have any direct couplings
to the MSSM fields, the leading effects on the
standard model will come from loop graphs involving the fields
$\Phi$ and $\tilde\Phi$, which depend on their mass \Eq{messmassex}.
These fields are therefore the messengers of SUSY breaking.
The messenger mass parameters are the only parameters in this model that
will have an observable effect on physics in the visible sector.

The only unsatisfactory feature of this model is that the mass terms
$M$ and $v^2$ are put in by hand.
There is a large literature on models that break SUSY dynamically,
in which the SUSY breaking messenger mass arises from dimensional
transmutation (see \Ref{GRGMSB} for a review).
However, since the messenger mass parameters are the only parameters
in the SUSY breaking sector that have observable effects in the visible
sector, we will be content to assume that a fully satisfactory model
exists and work out the consequences.

We therefore simply assume that there are charged messengers with
SUSY breaking mass
\beq
\De\scr{L}_{\rm mess} = \myint d^2\th\, \scr{M} \tilde{\Phi} \Phi
+ \hc
\eeq
with chiral superfield mass parameter
\beq
\scr{M} = M + \th^2 F.
\eeq
The messengers must be in a vector-like representation of the
standard model gauge group to allow this mass.
In order to keep the successful unification of couplings in the
MSSM, we can take the messengers in complete $SU(5)$ multiplets,
such as ${\bf 5} \oplus \bar{\bf 5}$.
After integrating out the auxiliary fields, the scalar masses are
\beq
\De\scr{L}_{\rm mess} \to -|M|^2 
(\phi^\dagger \phi + \tilde{\phi}^\dagger \tilde\phi)
+ (F \tilde\phi \phi + \hc).
\eeq
We can rephase $\phi$ and $\tilde\phi$ to make $F$ real, in which case
the scalar mass eigenstates are
$(\phi \pm \tilde{\phi}) / \sqrt{2}$,
with masses $|M|^2 \pm F$.
We see that stability of the vacuum $\avg{\phi} = \avg{\tilde\phi} = 0$
requires $F \le |M|^2$.
The fermion masses are unaffected by $F$, and so the fermion mass is $|M|$.

When we integrate out the messengers at loop level, the resulting low-energy
effective theory breaks SUSY.
At one loop, we get a gaugino mass from the diagram
\beq[gauginomassGMSBest]
m_{\rm gaugino} \sim \BoxedEPSF{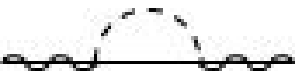}
\sim \frac{g^2}{16\pi^2} \frac{F}{M}.
\eeq
and at two loops, we get a scalar mass from diagrams like
\beq[scalarmassGMSBest]
m_{\rm scalar}^2 \sim
\BoxedEPSF{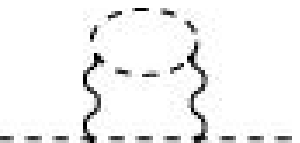} + \cdots
\sim \left( \frac{g^2}{16\pi^2} \right)^2 \left| \frac{F}{M} \right|^2.
\eeq
These are the leading terms in an expansion in powers of $F$.
(The corrections are suppressed by powers of $F/M^2$.)
The estimates give the right order of magnitude even if $F \sim M^2$.
One immediate consequence of this is that the masses of the scalars
and gauginos are of the same order, which is important for getting
a realistic and natural model of SUSY breaking.
Also, note that the scalar masses depend only on the gauge
quantum numbers of the scalars, and are therefore flavor-blind.
This gives a natural solution to the SUSY flavor problem.
Furthermore, the spectrum is determined by just a few
parameters, so this is a highly predictive framework for
SUSY breaking.
This mechanism is called `gauge mediation' of SUSY breaking,
for obvious reasons.
\bex
Show that the spurion $F/M$ has the right $U(1)_R$ charge to give
rise to a gaugino mass.
Use symmetry properties and dimensional analysis
to show that the leading contribution to
the scalar mass for small $F$ must be proportional to $|F / M|^2$.
Also, show that the subleading terms for small $F$ give rise to
fractional corrections of order $|F / M^2|^2$ to
\Eqs{gauginomassGMSBest} and \eq{scalarmassGMSBest}.
\eex

If we want the superpartner masses to of order $100\GeV$ or more, we
need to have $F/M \sim 5$--$50\TeV$.
However, the mass scale $M$ can be quite large, keeping this ratio
fixed.
An upper bound on $M$ is obtained by requiring that the gauge
mediated contributions to scalar masses to be sufficiently larger
than the contributions from Planck suppressed higher-dimension operators.
If SUSY is broken primordially by the VEV of a chiral superfield
\beq
\avg{X} = \th^2 F_0,
\eeq
then no symmetry can forbid operators of the form
\beq
\De\scr{L} \sim \myint d^4\th\, \frac{1}{\MP^2} X^\dagger X Q^\dagger Q.
\eeq
These will in general give rise to flavor-violating scalar masses of order
\beq
\De m_{\tilde Q}^2 \sim \frac{F_0^2}{\MP^2}.
\eeq
Demanding that $\De m_{\tilde Q}^2$ is small enough to avoid
FCNC's gives the bound
\beq
\sqrt{F_0} \lsim 10^{10}\GeV \left( \frac{m_{\tilde Q}}{500\GeV} \right)^{3/2}.
\eeq

Note that the primodial SUSY breaking scale $F_0$ need not be the same
as the scale $F$ in the messenger mass.
In particular, it is possible to have $F_0 \gg F$ if SUSY breaking is
communicated weakly to the messengers.
However, we must have $F_0 \gsim F$, so the largest possible value of
$M$ is obtained when $F_0 \sim F$:
\beq[F0bound]
M \lsim 10^{15} \GeV
\left( \frac{m_{\tilde Q}}{500\GeV} \right)^3.
\eeq

% --------------------------------------------------------------------------
\subsection{The Gauge Mediated Spectrum}
We now turn to the calculation of the gauge mediated spectrum.
Even at the qualitative level, it is crucial to know the
signs of the squark and slepton
mass-squared terms.
If any of these are negative, the theory does not have a minimum
that preserves color and electomagetism, which is certainly ruled out!

We will compute the induced SUSY breaking masses using
an elegant method due to Giudice and Rattazzi
\cite{GR} that makes essential use of superfield couplings.
We treat the messenger mass $M$ as a chiral superfield
\beq
\scr{M} = M + \th^2 F.
\eeq
This reduces the problem to how the superfield couplings in the
effective theory below the scale $M$ depend on $\scr{M}$.
The leading dependence for large $M$
is given by the RG, making the calculation
of the loop diagrams very simple.
For example, the value standard-model gauge coupling at a scale
$\mu < M$ can be obtained from the one-loop RG equation:
\beq[gaugerunex]
\frac{1}{g^2(\mu)} = \frac{1}{g'^2(\La)}
+ \frac{b'}{8\pi^2} \ln \frac{M}{\La}
+ \frac{b}{8\pi^2} \ln \frac{\mu}{M}.
\eeq
Here $g'$ is the gauge coupling in the theory above the scale $M$, and
$b'$ is the beta function coefficient in this theory,
while $g$ and $b$ are the corresponding quantities in the effective
theory below the scale $M$.
We started the running at an arbitrary scale $\La > M$.
For a non-abelian group,
\beq
b - b' = N,
\eeq
where $N$ is the number of messengers that get mass at the scale $M$
if the messengers are in the fundamental representation.
Note that $N$ is always positive.

The idea of Giudice and Rattazzi is to extend the formula
\Eq{gaugerunex} to superfield couplings.
We therefore have
\beq[taueff]
\tau(\mu) = \tau'(\La) + \frac{b'}{16 \pi^2} \ln\frac{\cal M}{\La}
+ \frac{b}{16 \pi^2} \ln\frac{\mu}{\cal M}.
\eeq
Here $\tau$ is the chiral superfield containing the (holomorphic)
gauge coupling.
Both sides of this equation are now well-defined
chiral superfields.
We can then compute the gaugino mass just by taking the higher $\th$
components of both sides:
\beq
\frac{m_{\la}(\mu)}{g^2(\mu)} &= -[ \tau(\mu) ]_{\th^2}
 = \frac{b - b'}{16\pi^2} [\ln \scr{M}]_{\th^2}
\nonumber\\
\eql{gauginoGRbasic}
&= \frac{N}{16\pi^2} \, \frac{F}{M}.
\eeq
Note that we have assumed that the only SUSY breaking is contained in
$\scr{M}$.
In particular, the couplings at the cutoff $\La$ have no higher
$\th$ components, which means that SUSY is unbroken in the fundamental
theory above the scale $M$.
In components, this would have been a finite one-loop
computation, but this method reduces it to a simple RG calculation.

Note that the result \Eq{gauginoGRbasic}
includes the running from the matching scale $M$
down to scales $\mu < M$.
We can find the value of the gaugino mass at the matching
scale $M$ by expanding about $\mu = \scr{M}$.
We illustrate this method here because it
is very useful for the scalar masses to be discussed below.
For the gaugino mass, we write
\beq
\tau(\mu) = \tau(\scr{M}) + \left. \frac{d \tau}{d\ln\mu} \right|_{\mu = \scr{M}}
\ln \frac{\mu}{\scr{M}} + \scr{O}\left( \ln^2 \frac{\mu}{\scr{M}} \right).
\eeq
When we take the $\th^2$ component of both sides, the terms of order
$\ln^2(\mu / \scr{M})$ do not contribute in the limit $\mu \to M$.
We then have
\beq
\lim_{\mu \to M}
[\tau(\mu)]_{\th^2} = [\tau(\scr{M})]_{\th^2}
+ \left. \frac{d \tau}{d\ln\mu} \right|_{\mu = M}
\left[ \ln \frac{\mu}{\scr{M}} \right]_{\th^2}.
\eeq
We then compute
\beq
{}[\tau(\scr{M})]_{\th^2} &= \frac{F}{M} \frac{\d \tau(M)}{\d\ln M}
= \frac{F}{M} \frac{\d \tau'(M)}{\d\ln M}
\nonumber\\
&= \frac{F}{M} \frac{b'}{8\pi^2}.
\eeq
Note that in our expansion the UV
couplings are held fixed, which is why the result is proportional to 
the beta function in the theory above the scale $M$.
Putting this together, we obtain
\beq
\lim_{\mu \to M}
[\tau(\mu)]_{\th^2} = \frac{b' - b}{8\pi^2} \frac{F}{M},
\eeq
in agreement with \Eq{gauginoGRbasic}.

This method is even more powerful when used to compute scalar masses.
These are extracted from the wavefunction coefficient $Z$ via
\beq
m^2 = -[\ln Z]_{\th^2 \bar{\th}^2}.
\eeq
Here $Z$ is a real superfield, so it depends on $\scr{M}$ via the
real superfield
\beq
\ln M \to \ln |\scr{M}| = \ln|M| + \sfrac 12
\left( \th^2 \frac{F}{M} + \hc \right).
\eeq
Expanding about $\mu = \scr{M}$, we have
\beq[Zththexpand]\bal
\lim_{\mu \to M} [ \ln Z(\mu) ]_{\th^2\bar{\th}^2}
&= [\ln Z(\scr{M})]_{\th^2 \bar{\th}^2}
+ \left([\ga(\scr{M})]_{\th^2}
\left[ \ln \frac{\mu}{\scr{M}} \right]_{\bar{\th}^2} + \hc \right)
\\
&\qquad
+ \frac 12 \frac{d\ga}{d\ln\mu}(M)
\left[ \ln^2\frac{\mu}{\scr{M}} \right]_{\th^2\bar{\th}^2},
\eal\eeq
where
\beq
\ga(\mu) = \frac{d\ln Z}{d\ln\mu}
\eeq
is the anomalous dimension in the effective theory below the scale $M$.
As in the calculation of the gaugino mass,
we must perform the expansion keeping the UV
cutoff fixed, which means that we must expand in $M$ in the fundamental theory.
We therefore have
\beq
\nonumber
{}[\ln Z(\scr{M})]_{\th^2\bar{\th}^2}
&= \frac 14 \left| \frac{F}{M} \right|^2
\left( \frac{\d}{\d\ln M} \right)^2 \ln Z'(M)
\\
&= \frac 14 \left| \frac{F}{M} \right|^2
\frac{d \ga'}{d \ln\mu}(M),
\eeq
where
\beq
\ga'(\mu) = \frac{d\ln Z'}{d\ln\mu}
\eeq
is the anomalous dimension in the theory above the scale $M$.
Similarly,
\beq
{}[\ga(M)]_{\th^2}
&= \frac 12 \frac FM \frac{\d}{\d\ln M} \ga(g'(M))
\nonumber\\
&= \frac 12 \frac FM \frac{\d\ga}{d g_i} (M) \be'_i(M),
\eeq
where $g_i$ ($g'_i$) denotes the dimensionless couplings of the theory
below (above) the scale $M$, and $\be_i$ ($\be'_i$) are the 
corresponding beta functions, \eg
\beq
\be_i = \frac{d g_i}{d\ln\mu}.
\eeq
Putting it all together, we obtain
\beq
m^2(M) &= -\!\!\! \lim_{\mu \to M} [ \ln Z(\mu) ]_{\th^2 \bar{\th}^2}
\nonumber\\
&= \frac 14 \left| \frac{F}{M} \right|^2 \left[
- \frac{\d\ga'}{\d g'_i} \be'_i
+ 2 \frac{\d\ga}{d g_i} \be_i'
- \frac{\d\ga}{\d g_i} \be_i
\right].
\eeq
Here all anomalous dimensions are evaluated at $\mu = M$.
This shows that the gauge mediated scalar mass at the threshold
is a simple function of the anomalous dimensions of the theory.
From this formula, we see that the scalar masses arise at two
loops, since both $\ga$ and $\be$ start at one loop.

Note that
we have performed a two-loop finite matching calculation using
only the RG equations.
We see that the threshold corrections are determined completely
by the anomalous dimensions and beta functions of the theory.
This is another illustration of the power of superfield couplings.

Squarks and sleptons do not couple directly to the messengers, so
they have $\ga' = \ga$ at one loop.
(This means that $\ga'$ is the same function of the couplings $g'$
as $\ga$ is of the couplings $g$.)
In this case, the expression for the scalar mass simplifies further:
\beq
m^2(M) = \frac{1}{4} \left|\frac{F}{M}\right|^2
\frac{\d\ga}{\d g_i} (\be'_i - \be_i).
%= \frac{1}{4} \left|\frac{F}{M}\right|^2
%\left( \frac{\d\ga'}{\d\ln\mu} - \frac{\d\ga}{\d\ln\mu} \right).
\eeq
The one-loop RG for a kinetic coefficient (of a quark field, say) 
from a gauge loop is
\beq
\mu\frac{d \ln Z}{d\mu} = \frac{c}{4\pi^2} g^2,
\eeq
where $c$ is the quadratic Casimir of the field.
For a fundamental representation of an $SU(N)$ gauge group,
$c = (N^2 - 1)/(2N)$.
Putting this in, we obtain
\beq[GMSBscalarform]
m^2(\mu = M) = \frac{g^4}{(16\pi^2)^2} 2 c N.
\eeq
Note that the scalar masses are positive at the matching scale
$\mu = |M|$, which is certainly a good starting point for a realistic
model.
RG evolution down to the weak scale can make the up-type Higgs mass
run negative (due to the large top Yukawa coupling), triggering
electroweak symmetry breaking.

Using the same techniques, we can see that
\beq
\lim_{\mu\to M} [\ln Z(\mu)]_{\th^2} = \frac 12 \frac{F}{M} (\ga' - \ga).
\eeq
Again, for particles that do not couple directly to the messengers
$\ga' = \ga$ at one loop, and so we do not get $A$ terms at one loop.
(This is also obvious from the fact that there are no one-loop diagrams
that could give an $A$ term.)
Direct couplings of the quarks and leptons to the messengers
violate flavor symmetries, but the Higgs can have nontrivial
couplings to the messengers.
Some of the consequences of this are explored in \Refs{GMSBmesscoup}.
%
%From \Eq{lnZonederiv} we 
%also see that the $B$ terms vanish at the scale $\mu = |M|$:
%\beq
%B(\mu = |M|) = [\ln Z(\mu = |M|)]_{\th^2} = 0.
%\eeq
%There are nonzero $A$ terms generated by running down to $\mu$,
%from gaugino loops, and therefore
%\beq
%A \sim \frac{g^2}{16\pi^2} m_\la \ln \frac{M}{M_W},
%\eeq
%which is negligible for most purposes.

It is important to remember that the results above are
only the leading result in an expansion in powers of
$F / M^2$.
In the effective theory below the messenger scale $M$, these
additional terms are parameterized by terms with additional SUSY
covariant derivatives, such as
\beq
\De \scr{L}_{\rm eff} \sim \myint d^4\th\,
\left| \frac{D^2 M}{M^2} \right|^2 Q^\dagger Q
\sim \left| \frac{F^2}{M^3} \right|^2 \tilde{Q}^\dagger \tilde{Q}
+ \cdots
\eeq
Unlike the leading terms computed above,
these terms are not related to the dimensionless couplings of
the low-energy theory, and therefore require an independent calculation.
This calculation has been performed in \Refs{fullloopGMSB}.
The result is that the scalar mass in particular is very insensitive to
corrections unless $F$ is very near $|M|^2$.

% ---------------------------------------------------------------------------
\subsection{Phenomenology of Gauge Mediation}
We now mention briefly some highlights of the phenomenology of
gauge-mediated SUSY breaking.
For more detail, see \Ref{GRreview} and references therein.

First, note that because superpartner masses are controlled by
gauge couplings, colored states will be much heavier than uncolored
states.
In particular, the ratio of stop masses to right-handed slepton
masses is of order
\beq
\frac{m_{\tilde{t}}}{m_{\tilde{e}_R}} \sim \sqrt{3}\, \frac{g_3^2}{g_1^2}
\sim 10.
\eeq
(The factor of $\sqrt{3}$ comes from the color factor
$c \sim 3$ in \Eq{GMSBscalarform}.)
The experimental bound $m_{\tilde{e}_R} \ge 99\GeV$ therefore implies
$m_{\tilde{t}} \gsim 980\GeV$, which implies a sizable fine-tuning for
electroweak symmetry breaking.
The actual value of $m_{\tilde t}$ can be smaller, but fine-tuning
is a concern in gauge mediated SUSY breaking.
As always, this fine-tuning is more severe if SUSY is broken at high
scales.

Another important feature of gauge-mediated SUSY breaking is that
the gravitino is generally the LSP.
In standard scenarios for SUSY breaking, the gravitino gets a mass
\beq
m_{3/2} = \frac{F_0}{\sqrt{3}\, \MP} \sim 100\GeV
\left( \frac{\sqrt{F_0}}{10^{10} \GeV} \right)^2,
\eeq
where $F_0$ is the primordial scale of SUSY breaking.
(We will review the origin of this formula below when we discuss
supergravity.)
The bound \Eq{F0bound} implies that $m_{3/2} \ll 100\GeV$
as long as $F_0$ is well below its maximum natural value,
as suggested by fine-tuning considerations.

If the gravitino is the LSP, then all SUSY particles eventually decay
to gravitinos and ordinary particles.
To understand these decays, we use the fact that the gravitino
mass can be thought of as arising from a `super Higgs mechanism',
in which a massless spin $\frac 32$ field (2 degrees of freedom)
`eats' the massless spin $\frac 12$ Goldstino field (2 degrees of freedom)
to make a massive spin $\frac 32$ field.
This is in direct analogy to a massive spin 1 particle, which
can be thought of as a massless spin 1 particle (2 degrees of freedom)
and an `eaten' massless spin 0 Nambu-Goldstone field
(1 degrees of freedom).
We will not give the details here, but the important point is that
the massless spin $\frac 32$ field is part of the supergravity
multiplet, and therefore couples to matter with strength suppressed
by powers of $1/\MP$.
On the other hand, the `eaten' Goldstino couples to matter with
strength determined by the priomordial SUSY breaking scale $F_0$.
Since $F_0 \ll \MP^2$, matter couples dominantly to the Goldstino
field, and we can ignore the spin $\frac 32$ gravitino field.

Low energy theorems analogous to those for ordinary broken symmetries
tell us that the coupling of the Goldstino field $\tilde{G}_\al$ couples
to matter via the supercurrent $J^{\mu \al}$:
\beq
\scr{L}_{\rm int} = -\frac{1}{F_0} J^{\mu \al} \d_\mu \tilde{G}_\al
+ \hc + \scr{O}(\tilde{G}^2),
\eeq
where the supercurrent is
\beq
J^\mu_\al = (\bar\psi_a \tilde{\si}^\mu \si^\nu)_\al \d_\nu \phi^a
- \frac{i}{4\sqrt{2}} (\la_A \tilde{\si}^\mu \si^{\nu})_\al
F_{\mu\nu A}.
\eeq
This can be used to compute the decays of the other superpartners into
Goldstinos.

Since $\sqrt{F_0}$ is much larger that superpartner masses,
the heavy superpartners will decay rapidly to the next-to-lightest
superpartner (NLSP), which will then decay more slowly into Goldstinos.
(We are assuming that $R$ parity or a similar symmetry prevents rapid
decays of the NLSP.)
The phenomenology therefore depends on the value of $F_0$ and the
identity of the NLSP.
If the NLSP is Bino, its dominant decay is
\beq
\Ga(\chi_1^0 \to \ga \tilde{G}) \sim 10^{-3} \eV
\left( \frac{m_{\chi_1^0}}{100\GeV} \right)^5
\left( \frac{\sqrt{F_0}}{100 \TeV} \right)^{-4}.
\eeq
If the NLSP is the right-handed stau (the lightest of the right-handed
sleptons because of mixing effects), its dominant decay is
\beq
\Ga(\tilde{\tau}_R \to \tau \tilde{G}) \sim 10^{-3} \eV
\left( \frac{m_{\tilde{\tau}_R}}{100\GeV} \right)^5
\left( \frac{\sqrt{F_0}}{100 \TeV} \right)^{-4}.
\eeq
Depending on the value of $F_0$, the NLSP can have a visible
decay length:
\beq
L = \frac{c \ga}{\Ga}
\sim 10^{-2} {\rm cm}
\left( \frac{m}{100\GeV} \right)^{-5}
\left( \frac{\sqrt{F_0}}{100\TeV} \right)^4
\times \sqrt{E^2 / m^2 - 1}.
\eeq
For $\sqrt{F_0} \lsim 10^{-6}\GeV$ this decay is inside the detector.
Even for $\sqrt{F_0} \sim 100\TeV$ (the smallest allowed value)
this gives a displaced vertex small enough to be seen in a silicon
vertex detector.
Measurement of the decay length of the NLSP therefore gives direct
information about the scale of primordial SUSY breaking!

\subsection{Gravitino Cosmology}
We now make some brief remarks on gravitino cosmology in gauge-mediated
models with $R$ parity.
The gravitino is stable in these models,
and can therefore contribute to the
energy density of the universe today.
If the gravitino has a thermal abundance early in the universe,
it freezes out at temperatures of order its mass.
The relic abundance today is of order 
\beq
\Om_{3/2} \sim \left( \frac{\sqrt{F_0}}{10^6 \GeV} \right)^{-1}
\eeq
where $\Om$ is the fraction of critical density contributed by the
gravitino.
(See \Ref{KolbTurner} for a discussion
of this standard calculation.)
In order to avoid overclosing the universe we need $\Om_{3/2} \lsim 1$,
or $\sqrt{F_0} \lsim 10^6 \GeV$.
Note that this implies that the gravitino decays inside the detector
in collider experiments!
It is still possible to have $\sqrt{F_0} > 10^6 \GeV$ if the primordial
gravitino abundance is diluted by inflation with a low reheat temperature
or by significant late-time entropy production.

% ------------------------------------------------------------------
\section{`Need-to-know' Supergravity}
% ------------------------------------------------------------------
We now switch gears to a more formal subject: supergravity
(SUGRA).
SUGRA is the supersymmetric generalization of Einstein gravity,
and as such unquestionably has a fundamental place in a supersymmetric
world.
However, the gravitational force
is so weak that it is generally unimportant for
particle physics experiments, so we start by explaining the motivation
for a particle phenomenologist to learn about SUGRA.

One reason has already emerged in our discussion of gauge-mediated
SUSY breaking.
Namely, the superpartner of the spin-2 graviton is a the spin-$\frac 32$
gravitino.
Its interactions with ordinary matter are suppressed by powers of the
Planck scale, but it can have interesting (or dangerous)
cosmological effects.

Another motivation is the cosmological constant problem, which is
clearly a gravitational effect.
The cosmological constant can be naturally zero in the limit of
unbroken SUSY, but the cosmological constant problem comes back
when SUSY is broken.
There is at present no convincing solution to the cosmological
constant problem, but SUSY is the only known symmetry that can
explain why the cosmological constant smaller than the Planck scale,
and may therefore play a role in the eventual solution.

However, the primary motivation for studying SUGRA in these lectures
is that SUGRA can be the messenger of SUSY breaking.
In section \ref{sec:hidden} we already considered
gravity as the messenger of SUSY breaking.
However, we really only considered the effect of integrating out
heavy physics at the scale $\MP$.
We did not include the effects of the SUGRA fields, which are light!
In particular, we want to explore the idea that SUSY breaking
can be communicated to the visible
sector via the VEV of an auxiliary field in the SUGRA multiplet.
This auxiliary field is in some ways analogous to a $D$ field in SUSY
gauge theory:
if $\avg{D} \ne 0$, it will give rise to SUSY breaking for fields
that are charged under the gauge group.%
\footnote{One might wonder why we don't try to couple an additional $U(1)$
to the MSSM and break SUSY by $\avg{D} \ne 0$.
One obstacle to building a model of this kind is that the scalar masses are
proportional to $U(1)$ charges, which must occur with both signs in
order to cancel gauge anomalies.
Nonetheless, this approach may be viable \cite{DU1}.}
Since gravity couples universally, all fields are `charged' under SUGRA,
and we might expect that this gives rise to flavor-blind SUSY breaking.
This idea can be made to work, but it turns out to be rather
subtle and we will have to develop some formal machinery before we
can get to the physics.
Let us begin.

There are several different formalisms for SUGRA,
all of which are related by field redefinitions and
give equivalent physical results.
The simplest formulation for our purposes is the tensor calculus
approach.
The basic idea of this approach is to write off-shell supermultiplets
as a collection of component fields, and to define the usual superfield
operations directly on this collection of components.
For example, a chiral multiplet is written as
\beq
\Phi = ( \phi,\ \psi_\al,\ F ),
\eeq
and products of chiral superfields are defined by
\beq
\Phi_1 \Phi_2 = (\phi_1 \phi_2,\ \phi_1 \psi_{2 \al} + \phi_2 \psi_{1 \al},\ 
\phi_1 F_2 + \phi_2 F_1 + \psi_1 \psi_2).
\eeq
SUSY invariants are defined by taking the highest components of
superfields, \eg
\beq
\myint d^2 \th\, \Phi = F.
\eeq
For chiral and real superfields without SUGRA,
this is just a rewriting of the usual
rules for combining superfields.
In the tensor calculus approach, matter and gauge supermultiplets are
coupled to SUGRA by `covariantizing' the rules for combining superfields
and forming SUSY invariants.
The minimal off-shell SUGRA multiplet is
\beq
(e_\mu{}^a,\ \psi_{\mu\al},\ B_\mu,\ F_\phi),
\eeq
where $e_\mu{}^a$ is the 4-bein, $\psi_{\mu\al}$ is the gravitino field,
and $B_\mu$ and $F_\phi$ are vector and scalar auxiliary fields, respectively.
In the tensor calculus approach, the SUGRA fields are included by suitably
covariantizing the usual rules for multiplying supermultiplets and taking
their highest components to define SUSY invariants.
In particular, the 4-bein $e_\mu{}^a$ is
coupled according to the standard rules from
general relativity.%
\footnote{For a clear introduction to the 4-bein formalism of general relativity,
see \Ref{WeinbergGravity} or \Ref{GSW2}.}

For SUSY breaking, interested in
$\avg{F_\phi} \ne 0$, since a VEV for $B_\mu$ would break Lorentz invariance.
The dependence on $F_\phi$ is governed by supercovariance, and is closely
related to a local (gauged) conformal invariance of the theory.

To understand this local conformal invariance, let us see how it can be
introduced in Einstein gravity without SUSY.
There we can write a theory in a way that is invariant
under local scale transformations by introducing an additional real scalar.
The additional gauge symmetry can be used to gauge away the scalar,
so this theory is equivalent to ordinary Einstein gravity.
We first introduce local scale transformations acting on the metric as
\beq
g_{\mu\nu}(x) \mapsto \Om^2(x) g_{\mu\nu}(x).
\eeq
It is easy to see that the usual Einstein kinetic term is not invariant:
\beq
\Ga^\rho_{\mu\nu} \sim g^{-1} \d g
&\ \ \Rightarrow\ \ 
\Ga^\rho_{\mu\nu} \mapsto
\Ga^\rho_{\mu\nu} + \scr{O}(\d \Om),
\\
R_{\mu\nu} \sim \d\Ga + \Ga^2
&\ \ \Rightarrow\ \ 
R_{\mu\nu} \mapsto R_{\mu\nu} + \scr{O}(\d\Om),
\eeq
where $\scr{O}(\d\Om)$ denotes terms with derivatives acting on $\Om$.
Thererfore,
\beq
R = g^{\mu\nu} R_{\mu\nu} &\mapsto \Om^{-2} R + \scr{O}(\d\Om),
\\
\sqrt{-g} &\mapsto \Om^4 \sqrt{-g},
\eeq
and we see that the Einstein Lagrangian $\sqrt{-g}\, R$ is not invariant.
(Alternatively, it is clear that the Einstein Lagrangian is not conformally
invariant because it has a dimensionful
coefficient proportional to $\MP^2$.)
There is a 4-derivative action that is invariant under local scale
transformations, but there is no obvious way to make sense out of theories
whose leading kinetic term has 4 derivatives.

To make the Lagrangian invariant, we introduce a
real scalar $\eta$ 
transforming under local scale transformations as
\beq
\eta(x) \mapsto \Om^{-2}(x) \eta(x).
\eeq
We can then write an invariant action in terms of the invariant `metric'
\beq
\tilde{g}_{\mu\nu} = \eta\, g_{\mu\nu}:
\eeq
\beq[EinsteinLconformal]
\myint d^4 x \sqrt{-\tilde{g}}\, \tilde{R}
= \myint d^4 x \sqrt{-g} \left[ \eta^2 R - 6 (\d\eta)^2 \right].
\eeq
The signs are such that if the kinetic term for gravity has the right sign
for positive energy,
the kinetic term for the scalar has the `wrong' sign (negative energy).
Usually, a `wrong' sign kinetic term means that the theory has a catastrophic
instability to creation of negative energy modes.
However, this is not a disaster in this case, because $\eta$ is not a physical
degree of freedom: it can be gauged away.
In fact, this theory is equivalent to Einstein gravity, as we can
easily see by chosing the gauge
\beq[GRgauge]
\eta(x) \to \MP.
\eeq
\Eq{GRgauge} is a good gauge choice as long as $\eta$ is everywhere nonzero,
which is good enough for perturbative expansions.%
\footnote{It may be disturbing that the strength of the gravitational coupling is
apparently determined by a gauge choice.
However, one must remember that only ratios of scales have physical significance.
An operator of dimension $d$ in the lagragian will be proportional to $\eta^{4 - d}$\
by scale invariance, so the value of $\eta$ just sets the overall scale.}

We have seen that we can rewrite Einstein gravity as a theory with an extra
gauge symmetry (local scale invariance) and an extra scalar field.
In fact, scale invariance implies invariance under an extended set of
symmetries, the so-called conformal symmetries.%
\footnote{For an introduction, see \Ref{Ginsparg}.}
The scalar field $\eta$ is called the `conformal compensator.'
%We can choose a gauge where the scalar field takes a constant value, which sets
%the scale for gravity.

The same trick is useful in writing the SUGRA Lagrangian.
This approach is called the superconformal tensor calculus.
The full group of symmetries is the superconformal tranformations, which includes
scale transformations and a $U(1)_R$ symmetry.%
\footnote{%
The appearance of local scale symmetry can be understood in a deductive way
in the superfield formulation of SUGRA.
For a discussion of superfield SUGRA at the level of these
lectures, see \Refs{LLP}.}
One writes a theory that is invariant under local superconformal transformations,
based on the superconformal supergravity multiplet %(quite a mouthful, isn't it?)
\beq
(e_\mu{}^a,\ \psi_{\mu a},\ B_\mu,\ R_\mu).
\eeq
Here $B_\mu$ and $R_\mu$ are vector auxiliary fields.
To break the superconformal symmetry dowm to super-Poincar\'e symmetry,
one introduces a superconformal 
compensator supermultiplet, which is a chiral multiplet
\beq
\phi = (\eta,\ \chi,\ F_\phi).
\eeq
The real part of the scalar complex field $\eta$ plays the same role as the
real scalar field called $\eta$ above.
The dimension and $R$ charge of are
\beq
d(\phi) = 1,
\qquad
R(\phi) = \sfrac 23.
\eeq
Note the dimension is the same as the real scalar $\phi$ of \Eq{EinsteinLconformal},
and the $R$ charge is such that the superpotential $\int d^2\th \phi^3$ is
$U(1)_R$ invariant.
To see that the theory with the compensator is equivalent to ordinary
non-conformal SUGRA, one makes the gauge choice
\beq
\phi \to (1,\ 0,\ F_\phi),
\qquad
R_\mu \to 0.
\eeq
This discussion has been very sketchy.
For more details, see \Ref{Cremmer}.

The utility of all this formalism is that the couplings of the
superfield $\phi$,
and hence $F_\phi$, are completely fixed by superconformal invariance.
To determine the couplings of $\phi$ it is sufficient to keep
track of scale transformations and $U(1)_R$ transformations,
which are determined by the dimension $d$ and the $R$ charge.
The basic rule is that the Lagrangian $\scr{L}$ has $d(\scr{L}) = 4$
and $R(\scr{L}) = 0$.
For a SUSY Lagrangian of the form
\beq[stdLagSUSYAMSB]
\scr{L} = \myint d^4\th\, f + \left( \myint d^2\th\, W + \hc \right)
\eeq
this means that
\beq
d(f) &= 2,
\qquad
\ \,R(f) = 0,
\\
d(W) &= 3,
\qquad
R(W) = +2.
\eeq
It is convenient to choose all chiral and vector matter multiplets
to have $d = 0$ and $R = 0$.
(It may appear strange to choose $d = 0$ for matter fields,
but we will see that we can make field redefinitions so that $d$
coincides with the usual mass dimension.)
For a Lagrangian of the form \Eq{stdLagSUSYAMSB}, this implies
in particular $d(f) = 0$, so it is not superconformally invariant.
To make it invaraint, we use the conformal compensator.
%\beq
%\scr{L} = \myint d^4\th\, \phi^\dagger \phi f
%+ \left( \myint d^2 \th\, \phi^3 W + \hc \right),
%\eeq
%where $f$ is real and $W$ is chiral.
To convariantized kinetic and superpotential terms
for a Lagrangian of the form \Eq{stdLagSUSYAMSB} are then
\beq[KahlerwSUGRA]
\myint d^4\th\, \phi^\dagger \phi\, f
&= e \, \Bigl[ f |_{\th^4} + ( f |_{\th^2} \cdot F_\phi^\dagger + \hc )
\nonumber\\
& \qquad\quad
+ f| \cdot ( F_\phi^\dagger F_\phi + 6 R(g) + \bar\psi i \sla\partial \psi )
+ \hbox{\rm fermions}
 \Bigr],
\\
\eql{WwSUGRA}
\myint d^2\th\, \phi^3\, W &= e \, \Bigl[
W|_{\th^2} 
+ W | \cdot (3 F_\phi + \psi \psi)
+ \hbox{\rm fermions}
\Bigr],
\eeq
where $e = \det(e_\mu{}^a)$ is the determinant of the 4-bein and
the terms involving the gravitino $\psi$ have been written only
schematically.
Note that the constant term in $f$ contains a kinetic term for
the 4-bein and gravitino field, and a constant term in $W$ contains
a mass term for the gravitino.
We have not written terms involving the matter fermions, since we are
interested in SUSY breaking.

For a gauge field with field strength $W_\al$, note that
$d(W^\al W_\al) = 3$ and $R(W^\al W_\al) = 2$, so there is no
$\phi$ dependence in the standard gauge kinetic terms:
\beq[GaugewSUGRA]
\De\scr{L} = \myint d^2\th\, S(Q) W^\al W_\al + \hc
\eeq
The sum of \Eqs{KahlerwSUGRA}, \eq{WwSUGRA}, and \eq{GaugewSUGRA}
is the most general Lagrangian terms with 2 or fewer derivatives,
coupled to supergravity.

Now we are (finally) ready to make our first main point.
Consider a theory with no dimensionful couplings.
The Lagrangian can be written schematically as
\beq
\scr{L} = \myint d^4\th\, \phi^\dagger \phi \, Q^\dagger e^V Q
+ \left[ \myint d^2\th \left( \frac{1}{g^2} W^\al W_\al\
+ \phi^3 Q^3 \right) + \hc \right].
\eeq
It appears that this has nontrivial couplings between $\phi$
and the matter fields, the field redefinition
\beq[hatteddef]
\hat{Q} = \phi\, Q
\eeq
removes the $\phi$ dependence completely:
\beq
\scr{L} = \myint d^4\th\, \hat{Q}^\dagger e^V \hat{Q}
+ \left[ \myint d^2\th \left( \frac{1}{g^2} W^\al W_\al\
+ \hat{Q}^3 \right) + \hc \right].
\eeq
Note that $\phi$ and $Q$ are both chiral superfields, so $\hat{Q}$ is also
chiral.
Note also that this redefinition implies that the superconformal dimension of
$\hat{Q}$ coincides with its canonical dimension, \ie $d(\hat{Q}) = 1$.

In the case of the MSSM, the only dimensionful parameter is the
$\mu$ term.
Coupling this to SUGRA and using the canonical `hatted' fields
defined in \Eq{hatteddef}, the SUSY breaking part of the Lagrangian would be
(assuming $\avg{F_\phi} \ne 0$)
\beq
\De\scr{L}_{\rm SUSY\,break} = \mu \avg{F_\phi} H_u H_d + \hc
\eeq
That is, SUSY is broken only by a $B\mu$ term.
This does not give rise to a realistic model (\eg\ the squarks and
sleptons are much lighter than the Higgs).
However, we will see that there are important loop effects that can make
this form of SUSY breaking realistic.

The fact that $F_\phi$ decouples from a conformally invariant Lagrangian
at tree level would seem to imply that there are no supergravity corrections to the
potential at tree level.
However, this is not quite correct, because
the kinetic term for scalars includes a non-minimal
coupling to gravity:
\beq
\myint d^4\th\, \phi^\dagger \phi \, Q^\dagger Q
= e \left[ 
|\d Q|^2
+ \sfrac 16 |Q|^2 R(g) + \cdots \right].
\eeq
This coupling means that the scalar fields in general mix with gravity.
We can eliminate the non-standard scalar couplings by a field redefinition.
For a component Lagrangian of the form
\beq
\scr{L} = \sqrt{-g} \left[ ( \MP^2 + f(Q)) R(g) - V(Q) \right]
\eeq
we can redefine the metric
\beq
g_{\mu\nu} = \Om^2 \hat{g}_{\mu\nu},
\eeq
to obtain
\beq
\scr{L}  =\sqrt{-\hat{g}} \, \Om^4 \left[
\Om^{-2} ( \MP^2 + f(Q) ) R(\hat{g}) - V(Q) 
+ \scr{O}(\d f / \MP^2) \right],
\eeq
where the omitted terms involve derivatives acting on $f$.
Choosing
\beq
\Om^2 = \frac{\MP^2}{\MP^2 + f(Q)},
\eeq
we obtain
\beq
\scr{L} = \sqrt{-\hat{g}} \left[ \MP^2 R(\hat{g}) - \hat{V}(Q)
+ \scr{O}(\d f / \MP^2) \right],
\eeq
where
\beq
\hat{V}(Q) = \frac{V(Q)}{[1 + f(Q) / \MP^2]^2}.
\eeq
We have eliminated the non-standard couplings to gravity at the price
of multiplicatively changing the scalar potential.
This choice of definition of metric is often called `Einstein frame.'
The additional terms involving derivatives acting on $f$ mean that the
new Lagrangian does not contain canonically normalized scalar fields
in general.
Further field redefinitions of the scalar fields can make these
canonical.
When all this is done, the expression for the potential for the
canonically normalized scalar fields with no non-standard couplings
to gravity is more complicated.

If we go to Einstein frame in supergravity, the
connection to scale invariance is obscured.
For purposes of understanding SUSY breaking,
it is often better not to go to Einstein frame, as we will see.

% ----------------------------------------------------------------------
\subsection{\label{subsec:Polonyi}SUSY Breaking in SUGRA: Polonyi Model}
In this subsection and the following two, we consider SUSY breaking
in the presence of SUGRA.
Readers who are willing to take it for granted that $\avg{F_\phi}$
will be nonzero in the presence of SUSY breaking can skip to section
\ref{sec:AMSB} below.

The Polonyi model is the simplest model of SUSY breaking, and we now
consider what happens when it is coupled to SUGRA.
The Lagrangian is
\beq
\nonumber
\scr{L} &= \myint d^4\th\, \phi^\dagger \phi
[ -3 \MP^2 + f(X, X^\dagger) ]
\\
\eql{PolonyiwSUGRA}
&\qquad 
+ \left( \myint d^2\th\, \phi^3 [ c + \ka X ] + \hc \right).
\eeq
If the dimensionful parameters $c$ and $\ka$ are small in units of $\MP$
we expect the SUGRA corrections to the scalar potential to be small
perturbations.
In the absence of gravity, the scalar potential is
\beq
V = \frac{|\ka|^2}{f_{X^\dagger X}},
\eeq
where $f_{X^\dagger X} = \d^2 f / \d X^\dagger \d X$.
We assume that the \Kahler function $f$
is nontrivial so that this potential has a nontrivial minimum in which
SUSY is broken.
Since the superpotential is the most general linear function of $X$,
we can shift the field so the minimum occurs at $\avg{X} = 0$.

To find the SUGRA corrections, we write out the terms without derivatives
in the Lagrangian
\beq
\scr{L} &= F_\phi^\dagger F_\phi (-3 \MP^2 + f)
+ (F_X F_\phi^\dagger f_X + \hc)
\\
\nonumber
&\qquad
+ f_{X^\dagger X} F_X^\dagger F_X
\\
\nonumber
&\qquad
+ 3 F_\phi (c + \ka X) + \hc
\\
&\qquad
+ \ka F_X + \hc + \cdots
\eeq
%where $f_X = \d f / \d X$, \etc
%In the second equation, we have written all terms without derivatives,
%which are all we need to find the scalar potential.
%Note that the first term in \Eq{PolonyiwSUGRA} is the kinetic term for gravity.
%If the dimensionful parameters $c$ and $\ka$ are small in units of $\MP$
%we expect the SUGRA corrections to the scalar potential to be small
%perturbations.
%To see this, let us first integrate out the auxilary field $F_X$ to
%obtain
Integrating out $F_X$, we obtain
\beq
\nonumber
\scr{L} &\to -\frac{|\ka|^2}{f_{X^\dagger X}}
+ \left[ -3 \MP^2 + f - \frac{|f_X|^2}{f_{X^\dagger X}} \right]
F^\dagger_\phi F_\phi
\\
&\qquad + \left[ 3(c + \ka X) - \frac{\ka f_{X^\dagger}}{f_{X^\dagger X}}
\right] F_\phi + \hc
\eeq
If the dimensionful couplings in the Polonyi sector are small compared
to $\MP$, we can approximate the
coefficient of $F_\phi^\dagger F_\phi$ by $-3\MP^2$
and write
\beq[SUGRApotPolonyi]
V = \frac{|\ka|^2}{f_{X^\dagger X}}
- \frac{3|c|^2}{\MP^2} \left|
1 + \frac{\ka}{c} \left(X - \frac{f_{X^\dagger}}{3 f_{X^\dagger X}} \right)
\right|^2.
\eeq
Note that the SUGRA corrections to the potential are negative definite.
(This is related to the `wrong-sign' kinetic term for the conformal
compensator.)
This is a crucial property that allows us to cancel the positive
vacuum energy due to SUSY breaking and obtain a ground state with
vanishing cosmological constant.
This requires that we tune
\beq
|c|^2 \simeq \frac{\MP^2 |\ka|^2}{3 \avg{f_{X^\dagger X}}}.
\eeq
(Note that since $c \propto \MP$, it is a good approximation to
neglect the terms proportional to
$\ka/c$ in \Eq{SUGRApotPolonyi}.)
Because the \Kahler function $f$ is renormalized, this is not stable under
radiative corrections.
Note that if the fluctuations of $X$ about $\avg{X} = 0$ are
canonically normalized, then $\avg{f_{X^\dagger X}}\sim 1$
and $c \sim \ka \MP$.

In this vacuum, SUSY is broken by the auxiliary fields
\beq
\avg{F_X} &= -\frac{\ka^\dagger}{\avg{f_{X^\dagger X}}},
\\
\avg{F_\phi} &= \frac{c^\dagger}{\MP^2} \sim
\frac{\avg{F_X}}{\MP}
\eeq
up to corrections suppressed by powers of $1/\MP$,
and where we have assumed $\avg{f_{X^\dagger X}}\sim 1$ in the
last relation.
If we include the gravitino couplings, we find that there is also
a gravitino
mass
\beq
m_{3/2} = \frac{\avg{W}}{\MP}
= \frac{c}{\MP} = \avg{F_\phi^\dagger}.
\eeq
This is the origin of the formulas for the gravitino mass used in the
section on gauge mediated SUSY breaking.

We see that if SUSY is broken below the Planck scale
by the usual O'Raifeartaigh (or Polonyi) mechanism,
the only effect of SUGRA is to allow the fine-tuning of the cosmological
constant, and to generate a nonzero VEV for the auxiliary field
$F_\phi$.

% ----------------------------------------------------------------------
\subsection{\label{subsec:noscale}`No Scale' SUSY Breaking}
The above analysis assumed that SUSY is broken in the absence of SUGRA.
There is another possibility for SUSY breaking that can only occur in the
presence of SUGRA: `no scale' SUSY breaking.
To see how this works, consider a model of a single chiral superfield
$T$ with Lagrangian
\beq[noscaleL]
\scr{L} = \myint d^4\th\, \phi^\dagger \phi (T + T^\dagger)
+ \left( \myint d^2\th\, \phi^3 c + \hc \right),
\eeq
where $c$ is a constant.
Note that $T$ has been chosen to have dimensions of mass-squared.
This choice of Lagrangian is rather arbitrary, and in fact it is not
radiatively stable.
We will address this below, but let us start by understanding this
simple Lagrangian.

First, note that in the absence of gravity, the \Kahler term would be a
total derivative.
The field $T$ acquires a kinetic term only by mixing with gravity,
so this is only a healthy theory in the presence of gravity.
The terms with no derivatives are
\beq
\nonumber
\scr{L} &= F_\phi^\dagger F_\phi (T + T^\dagger)
+ (F_\phi^\dagger F_T + \hc)
\\
& \qquad + 3 F_\phi c + \hc + \cdots
\eeq
Note that we have not included the Kinetic term for gravity, since this
can be absorbed into a shift of $T$.
In order to get the right strength for gravity, we need
\beq[TVEV]
\avg{T} = -\sfrac 32 \MP^2.
\eeq
Varying with respect to $F_T$ tells us that $F_\phi = 0$,
and hence the potential vanishes identically.
In particular, this means that the cosmological constant vanishes.
On the other hand, the gravitino mass is nonzero:
\beq
m_{3/2} = \frac{\avg{W}}{\MP} = \frac{c}{\MP}.
\eeq
Thus, SUSY is broken with vanishing cosmological constant!
This kind of SUSY breaking is called `no scale' SUGRA for historical
reasons.
However, the fact that the potential vanishes identically also means
that the scalar field $T$ is completely undetermined.
Also, the form of the Lagrangian \Eq{noscaleL} is not preserved by
radiative corrections.

Suppose therefore that we add a nontrivial \Kahler corrections to the
Lagrangian above:
\beq
\De\scr{L} = \myint d^4\th\, \phi^\dagger \phi\, \De f(T, T^\dagger).
\eeq
Such corrections will in any case be induced radiatively, and may play
a role in the stabilization of $T$.
Let us treat $\De f$ perturbatively, and ask what are the conditions
that we get a vacuum that is `close' to the one found above.
It is easy to work out that the scalar potential to first order
in $\De f$:
\beq
\De V = -|c|^2 \De f_{T^\dagger T}.
\eeq
Therefore, if $f_{T^\dagger T}$ has a local maximum, the theory will
have a stable minimum where $\avg{T}$ is given by \Eq{TVEV}, as required.
In this vacuum, we have (to first order in $\De f$)
\beq
\avg{F_T} &= -3c^\dagger,
\\
\avg{F_\phi} &= -\left\langle
\frac{\De f_{T^\dagger T} F_T}{T + T^\dagger}\right\rangle
= -\frac{c^\dagger \avg{\De f_{T^\dagger T}}}{3\MP^2}
\eeq
Note if $\De f$ is small, we can make $\avg{F_\phi}$ as small as we want.
This kind of vacuum can be thought of as `almost no scale' SUGRA.
For a more complete discussion see \Ref{almostnoscale}.

% -----------------------------------------------------------------------
\subsection{The SUGRA Potential}
We now consider the SUGRA corrections to the scalar potential.
This is important to make contact between the approach to SUGRA
taken here, which emphasizes the auxiliary fields, and more conventional
treatments which use the SUGRA corrections to the potential as a starting
point.

We consider a 2-derivative Lagrangian of the form
\beq
\scr{L} = \myint d^4\th\, \phi^\dagger \phi f
+ \left( \myint d^2\th\, \phi^3 \, W + \hc \right),
\eeq
where $f$ and $W$ are functions of some matter fields $Q^a$.
In order to get the right kinetic term for gravity, we require
\beq
\avg{f} = -3\MP^2.
\eeq
The terms in the Lagrangian with no derivatives are
\beq
\scr{L} &= F^\dagger_\phi F_\phi f
+ (F_\phi^\dagger f_a F^a + \hc)
+ f_a{}^b F_b^\dagger F_a
\nonumber
\\
&\qquad
+ 3 F_\phi W + W_a F^a + \hc,
\eeq
where $f_a = \d f/ \d Q^a$, $f^a = \d f/ \d Q^\dagger_a$, \etc
Solving for the auxiliary fields, we find
\beq
F_a^\dagger &= -(\tilde{f}^{-1})_a{}^b \left[ W_b
- \frac{3 f_b}{f} W \right],
\\
F_\phi^\dagger &= -\frac{1}{f} (3 W + f^a F_a^\dagger),
\eeq
where $(\tilde{f}^{-1})_a{}^b$ is the matrix inverse of
\beq
\tilde{f}_a{}^b = f_a{}^b - \frac{1}{f} f_a f^b.
\eeq
Integrating out the auxiliary fields, we find after some algebra
\beq
V = (\tilde{f})_b{}^a \left( W_a - \frac{3W}{f} f_a \right)
\left( W^{\dagger b} - \frac{3 W^\dagger}{f} f^b \right)
- \frac{3 |W|^2}{f}.
\eeq
As discussed above, this is not the potential in Einstein frame.
To make the gravity kinetic term canoncial, we define the
Einstein frame metric
\beq
\hat{g}_{\mu\nu} = -\frac{f}{3\MP^2} g_{\mu\nu}.
\eeq
The potential in Einstein frame is then
\beq[SUGRAcanV]
\hat{V} = \left(\frac{3\MP^2}{f} \right)^2 V.
\eeq
Note also that the function $f$ is not what is called the
\Kahler potential in the SUGRA literature.
The \Kahler potnetial is related to $f$ by
\beq
f = -3\MP^2 e^{-K/3\MP^2}.
\eeq
With these relations, \Eq{SUGRAcanV} reduces to the standard
expression for the SUGRA potential (see \eg \Ref{WB}).

Let us apply these results to find the conditions for a vacuum
that \emph{preserves} SUSY and has a vanishing cosmological constant.
To preserve SUSY it is sufficient for all auxiliary fields to vanish
in the vacuum.
The condition $\avg{F_a^\dagger} = 0$ is equivalent to 
$W_a - 3 W f_a / f = 0$ provided that
$\avg{\tilde{f}_a{}^b}$ is a non-singular matrix.
This in turn is equivalent to the condition that
$W / f^3$ is stationary.
The condition $\avg{F_\phi^\dagger} = 0$ then imposes the
additional requirement that $\avg{W} = 0$.
Combining these, we see that the conditions for a SUSY vacuum
with vanishing cosmological constant are
\beq
W\ \hbox{\rm stationary},
\quad
\avg{W} = 0.
\eeq
Note that we can always impose $\avg{W} = 0$ by adding a constant
to the superpotential.
This shows that the SUSY-preserving vacua in the presence of
SUGRA with vanishing cosmological constant are in one-to-one
correspondence with the SUSY vacua in the absence of SUGRA.
In particular, the SUGRA corrections to the potential cannot
turn a SUSY preserving vacuum into a SUSY breaking one.

% -----------------------------------------------------------------------
\section{\label{sec:AMSB}Anomaly Mediated SUSY Breaking}
Now we finally have enough machinery to discuss SUGRA as the messenger
of SUSY breaking.
We therefore assume that the only source of SUSY breaking comes from a
non-vanishing value of $\avg{F_\phi}$.
This can be viewed as a SUGRA background in which we are calculating.
Let us consider a SUSY model with no dimensionful couplings in the
SUSY limit.
(The NMSSM is such a model, as is the MSSM if we omit the $\mu$ term.
We will see that the $\mu$ term in the MSSM will have to be treated
differently.)
As we have seen in the previous section, the fact that the
there are no dimensionful couplings means that at tree level
there is no SUSY breaking felt by the matter fields.
However, at loop level scale invariance is broken by the running of the
couplings.
We therefore expect SUSY breaking related to the conformal anomaly.
This is called `anomaly mediated SUSY breaking' (AMSB).
The original papers are \Refs{RS0,GLMR}.

To see how this works in a very concrete way, note that the
presence of UV divergences requires us to regulate the theory,
and the regulator necessarily introduces a mass scale, and therefore
breaks conformal symmetry.
For example, in a Wess-Zumino model, we can use higher-derivative
terms to regulate the theory:
\beq
\scr{L} = \myint d^4\th\, Z_0 \hat{Q}^\dagger
\left( 1 + \frac{\Box}{\La^2 \phi^\dagger \phi} \right) \hat{Q}
+ \left( \myint d^2\th\, \frac{\la}{6} \, \hat{Q}^3 + \hc \right),
\eeq
where $\Box$ is the covariant second derivative operator.
We have directly written the Lagrangian in terms of the canonical
`hatted' fields
defined by \Eq{hatteddef}.
The factors of $\phi$ are required
because the operator $\Box$ has $d = 2$ and $R = 0$.
The $\Box / \La$ term modifies the propagator of the component fields
in $\hat{Q}$.
For example, setting $\avg{F_\phi} = 0$ for the moment,
the scalar propagator is modified
\beq
\frac{i}{p^2} \to \frac{i}{p^2 - p^4 / \La^2}.
\eeq
This makes loops of $\hat{Q}$ fields UV convergent.%
\footnote{This regulator also introduces a ghost, \ie a state
with wrong-sign kinetic term at $p^2 = \La^2$.
However, this decouples when we take the limit $\La \to \infty$
and does not cause any difficulties.}
Reintroducing $\avg{F_\phi} \ne 0$ introduces a small splitting between
the regulated scalars and fermions required by the coupling to
the SUGRA background, but the loop diagrams are still finite.
Note that we are only regulating loops of $\hat{Q}$ fields,
not SUGRA loops.
SUGRA loops are suppressed by additional powers
of $\MP$, and are therefore much smaller than loop effects from
dimensionless couplings.

We now compute standard model loop corrections in the SUGRA background.
We do this (once again) by treating the couplings as superfields.
At one loop, the $\hat{Q}^\dagger \hat{Q}$ term in the 1PI effective
action is logathimically divergent:
\beq[wavefuncnrenormoneloopphi]
\BoxedEPSF{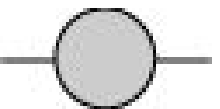} = Z_0 \left[
1 + \frac{\la^2}{26\pi^2} \ln \frac{\La |\phi|}{\mu}
+ \hbox{\rm finite}\right],
\eeq
where $\la$ is the physical Yukawa coupling.
Note the $\phi$ dependence in \Eq{wavefuncnrenormoneloopphi}
is required by conformal invariance.
There is no $\phi$ dependence in the `finite' part of the
amplitude because it is by definition independent of $\La$
as $\La \to \infty$.
The divergence ($\La$ dependence) must be cancelled by absorbing
it into the `bare' coupling $Z_0$, but we cannot absorb the $\phi$
dependence into $Z_0$ because we want all SUSY
breaking to be due to the nontrivial SUGRA background.
(If the superfield $Z_0$ had some nonzero SUSY breaking components,
this would be a theory with SUSY broken in the fundamental theory.)
This means that there is some SUSY breaking left over in the finite
part after we subtract the divergence.
We can view this as the replacement
\beq[musubamsb]
\ln\mu \to \ln\frac{\mu}{|\phi|}
= \ln\mu - \sfrac 12 (\th^2 F_\phi + \hc).
\eeq
The fact that this substitution parameterizes all the SUSY breaking
is clearly more general than than this example.
The cutoff $\La$ and the
renormalization scale $\mu$ always appear in the combination
$\La/\mu$, and for any real superfield coupling the correct
substitution is $\La \to \La|\phi|$, implying \Eq{musubamsb}.
Note also that \Eq{musubamsb} holds to all orders in perturbation theory.

We can therefore
compute SUSY breaking in real superfield couplings by
\beq
\frac{\d}{\d\th^2} = -\frac 12 F_\phi \frac{\d}{\d\ln\mu}.
\eeq
For example, the gaugino mass computed from the real superfield
gauge coupling is
\beq[gauginomassAMSB]
m_\la = -g^2 [R]_{\th^2} = \frac{\be(g)}{2g} F_\phi.
\eeq
We also have
\beq[BtermAMSB]
{}[\ln Z]_{\th^2} = -\sfrac 12 \ga F_\phi,
\eeq
which gives rise to nonzero $A$ terms.
Finally, we have soft masses
\beq[scalarmassAMSB]
m^2 = -{}[\ln Z]_{\th^2 \bar{\th}^2} = -\sfrac 14 |F_\phi|^2 \frac{d\ga}{d\ln\mu}.
\eeq
\Eqs{gauginomassAMSB}, \eq{BtermAMSB}, and \eq{scalarmassAMSB} are
exact in the sense that they hold to all orders in perturbation
theory.
They hold at each renormalization scale, and therefore define the
`AMSB renormalization group trajectory.'
As in gauge mediation, the dominant source of SUSY breaking comes
from gauge loops, and therefore the scalar masses are flavor-blind
and the model solves the SUSY flavor problem.
Also as in gauge mediation, gaugino masses arise at one loop and
scalar mass-squared parameters at two loops, so all SUSY breaking
masses are of the same order:
\beq
m_\la \sim \frac{g^2}{16\pi^2} \avg{F_\phi},
\qquad
m_0^2 \sim \frac{g^4}{(16\pi^2)^2} \avg{F_\phi}^2.
\eeq
To get SUSY breaking masses of order $100\GeV$, we need
$\avg{F_\phi} \sim 10\TeV$.

It is very interesting that the model gives the entire superpartner
spectrum in terms of a single new parameter $\avg{F_\phi}$,
which just sets the overall scale of the superpartner masses.
Let us check the crucial sign of the scalar masses.
For fields with only gauge interactions, we have
$\ga \sim +g^2$ and therefore
\beq
m^2 \sim -g \be_g |\avg{F_\phi}|^2.
\eeq
We see that if the gauge group is asymptotically free
($\be_g < 0$) the scalar mass-squared parameter is positive.
Unfortunately in the MSSM, the $SU(2)_W$ and $U(1)_Y$ gauge
groups are not asymptotically free, so sleptons have negative
mass-squared.
Adding more fields can only make this worse.
We cannot live on the AMSB renormalization trajectory.

Nonetheless, it is possible to have realistic SUSY breaking from
AMSB.
To understand this, let us consider the effect of massive
thresholds in AMSB.
Note that the formulas
\Eqs{gauginomassAMSB}, \eq{BtermAMSB}, and \eq{scalarmassAMSB}
are claimed to hold independently of the details of the high
energy theory, in particular the nature of ultrahigh energy
thresholds (\eg\ at the GUT scale).
Let us see how this works.

Consider some new chiral superfields $P$ and $\tilde P$ that
transform as a vectorlike representation of the standard model
gauge group, and which have a large supersymmetric mass term
\beq
\De \scr{L} = \myint d^2\th\, M \phi P \tilde{P} + \hc
\eeq
Note that we have included the superconformal compensator
in a normalization where the fields $P$ and $\tilde{P}$ have
canonical kinetic terms.
Because the fields $P$ and $\tilde{P}$ are charged, the gauge
beta functions will have different values above and below the 
scale $M$, so the SUSY breaking masses above and below the scale
$M$ are different.
To understand this, note that the $P$ threshold is not
supersymmetric because of the $\phi$ dependence.
Because of this, there is a \emph{gauge}-mediated threshold
correction at the scale $M$, with
\beq
\frac{F}{M} = F_\phi.
\eeq
For example, for gaugino masses, the threshold correction is
(see \Eq{gauginoGRbasic})
\beq
\De m_{\la} = \frac{\De \be_g}{2g} F_\phi.
\eeq
Adding this to the AMSB value above the threshold,
we find that the gaugino mass below the threshold is precisely
on the AMSB trajectory below the threshold.

Another way to understand this point is to consider again the superfield
couplings.
The holomorphic gauge coupling superfield $\tau$ below the threshold
$M$ is given by
\beq
\nonumber
\tau(\mu) &= \tau_0 + \frac{b'}{16\pi^2} \ln\frac{M}{\La}
+ \frac{b}{16\pi^2} \ln \frac{\mu}{M}
\\
&\to \tau_0 + \frac{b'}{16\pi^2} \ln\frac{M \phi}{\La \phi}
+ \frac{b}{16\pi^2} \ln \frac{\mu}{M \phi}.
\eeq
where $b$ and $b'$ are the beta function coefficients
below and above the scale $M$, respectively.
(The notation is the same as in section \ref{sec:GMSB}.)
We see that the $\phi$ dependence induced by $\La$ and $M$
exactly cancel in the contribution from above the scale $M$.
The gaugino mass below the scale $M$ is correctly given by
the substitution $\mu \to \mu / \phi$,
just as if the threshold did not exist.
We can think of $M$ as the new cutoff.

The fact that AMSB is independent of thresholds is very striking,
and makes the theory very predictive.
Unfortunately, we have seen that it is \emph{too} predictive,
and is ruled out by negative slepton mass-squared parameters.

This discussion however suggests a way out.
If a heavy threshold is not supersymmetric, the cancelation discussed
above no longer occurs, and we can be on a different RG trajectory
below the threshold.
If the $F/M$ of the threshold is much larger than $F_\phi$,
we have gauge mediation, and if it is much smaller it is irrelevant.
The only interesting case is where $F/M \sim F_\phi$, and we would
like this to occur naturally.

A very simple class of models where this occurs was first discussed
in \Ref{PR}.
Consider a theory with a singlet $S$ in addition to the vectorlike
fields $P$ and $\tilde{P}$, with superpotential terms
\beq
\De\scr{L} = \myint d^2\th \left[
\la S P \tilde{P} + \frac{S^n}{(M\phi)^{n-3}} \right] + \hc
\eeq
The potential for $S$ is
\beq
V = M^4 \left\{ n^2 \left| \frac{S}{M} \right|^{2(n-1)}
+ \left[ (n-3) \left( \frac{S}{M} \right)^n \frac{F_\phi}{M}
+ \hc \right] \right\}.
\eeq
Minimizing the potential we find
\beq[SVEVPR]
\left( \frac{\avg{S}}{M} \right)^{n-2} = \frac{n-3}{n(n-1)}
\frac{\avg{F_\phi}}{M}.
\eeq
From this we see that
\beq[FSVEVPR]
\frac{\avg{F_S}}{\avg{S}} = \frac{n-3}{n-1} \avg{F_\phi}.
\eeq
For $n > 3$ and $M \gg \avg{F_\phi}$, \Eq{SVEVPR} implies that
$\avg{F_\phi} \ll \avg{S} \ll M$,
while $\avg{F_S}/\avg{S} \sim \avg{F_\phi}$.
Because the coefficient in \Eq{FSVEVPR} is different from unity,
the theory will not be on the AMSB trajectory below the threshold.
These theories can be viewed as a combination of gauge and
anomaly mediation.
We can obtain a realistic SUSY breaking spectrum in this way
(in particular, the slepton masses can be positive).
See \Ref{PR} for more details.

There are also other ways proposed in the literature to make
AMSB realistic.
One class of models is similar to the proposal discussed above,
in that they have an additional `gauge mediated' contribution
to SUSY breaking that is naturally the same size as the AMSB
contribution \cite{PRtype}.
For a very different approach, see \Ref{otherAMSB}.

% -----------------------------------------------------------------------
\subsection{The $\mu$ Problem in Anomaly Mediation}
The $\mu$ problem is more severe in AMSB because we cannot simply
add a conventional $\mu$ term of the form
\beq
\De\scr{L} = \myint d^2\th\, \mu \phi H_u H_d + \hc
\eeq
The reason is that the explicit breaking of conformal invariance
means that
\beq
B\mu = \avg{F_\phi} \mu,
\eeq
which is far too large for $\avg{F_\phi} \sim 10\TeV$.

One possiblility is the NMSSM, discussed in
subsection \ref{subsec:NMSSM}.
This has no dimensionful couplings, and therefore this problem
is absent.
The effective $\mu$ term arises from a VEV for the singlet
$\avg{S}$, which is ultimately triggered by AMSB itself.
Given the fact that this 

Another possibility was pointed out in \Ref{RS0}.
If there is a chiral superfield $X$ with a shift symmetry
$X \mapsto X + \hbox{\rm constant}$,
then the $\mu$ term can arise from an operator of the form
\beq
\De\scr{L} &= \myint d^4\th\, \frac{\phi^\dagger}{\phi}
\frac{1}{M} (X + X^\dagger) H_u H_d + \hc
\eeq
Assuming $\avg{X} = 0$, we have
\beq
\De\scr{L} = \myint d^2 \th\, \frac{F_X^\dagger}{M} H_u H_d + \hc
\eeq
\ie we generate a $\mu$ term with no $B\mu$ term.
The $B\mu$ term can come from AMSB in this model.

Yet another possibility to generate the $\mu$ term from the
VEV of a singlet is described in \Ref{PR}.

% -----------------------------------------------------------------------
\subsection{Anomaly-Mediated Phenomenology}
Since the theory cannot be on the AMSB trajectory at low energies,
the low-energy phenomenology depends on how these problems are
resolved.
Discussions can be found in the original papers, quoted above.

We do want to point out, however, that these theories share the
fine-tuning problem of gauge mediated SUSY breaking, since
scalar masses arise from 2-loop gauge diagrams, and therefore
\beq
\frac{m_{\tilde{q}}^2}{m_{\tilde{e}_R}^2} \sim \frac{N_{\rm c} g_3^4}{g_1^4}.
\eeq

% -----------------------------------------------------------------------
\subsection{Naturalness of Anomaly Mediation}
So far we have not addressed the question of whether it is natural
for the theory to be on the AMSB trajectory.
What we would like is to have a theory that breaks
SUSY spontaneously in a hidden sector in
such a way that the breaking is communicated to the observable
sector dominantly through the SUGRA conformal compensator.

As we have seen in subsections \ref{subsec:Polonyi} and
\ref{subsec:noscale}, spontaneous SUSY breaking in SUGRA
generally gives
\beq
\avg{F_\phi} \lsim \frac{F_0}{\MP},
\eeq
where $F_0$ is the primordial SUSY breaking scale.
Generally, $F_0 = \avg{F_X}$, where $X$ is some chiral superfield.
In this case, we expect the effective theory to contain operators
of the form
\beq[scalarMPopagain]
\De\scr{L} \sim \myint d^4\th\, \frac{1}{\MP^2} X^\dagger X Q^\dagger Q
\eeq
where $Q$ are standard model fields.
As already discussed in section \ref{sec:hidden}, these
couplings have the quantum numbers of a product of kinetic
terms, and cannot be forbidden by any symmetries.
We therefore expect them to be present in any UV completion
of the theory at the Planck scale.
This gives rise to scalar masses of order
\beq
m^2 \sim \frac{\avg{F_X}^2}{\MP^2},
\eeq
which is much larger than the AMSB value.
(Furthermore, there is no reason for the term \Eq{scalarMPopagain}
to conserve flavor, so we expect the masses to give rise to
FCNC's.)
If we consider all the other possible
terms coupling the visible and the hidden
sector suppressed by powers of $\MP$
(see \Eq{hiddenSUSY}), we find that they can all naturally
be absent due to symmetries.
Therefore, the viability of anomaly mediation depends on
whether there are sensible models where the couplings
\Eq{scalarMPopagain} are naturally absent.

A simple rationale for this was given in \Ref{RS0}.
The idea is that the hidden and visible sectors are localized
on `branes' in extra dimensions.%
\footnote{%
We will not go into details here, but will just state the main
ideas.
For an introduction into many of the technical and conceptual
issues in theories with extra dimensions and branes, see the
lectures by Raman Sundrum at this school.}
That is, the standard model matter and gauge interactions are
localized on the visible brane, and the SUSY breaking sector
is localized on the hidden brane.
In fact, this type of setup naturally occurs in string theory,
\eg\ in the setup of \Ref{HoravaWitten}.
In the higher-dimensional theory, interactions like
\Eq{scalarMPopagain} are fobidden because $X$ and $Q$ are
localized on different branes, so the interaction is not
local.
Fields that propagate in the bulk can give rise to interactions
between $X$ and $Q$, so we must check whether interactions
like \Eq{scalarMPopagain} are generated in the $3+1$ dimensional
theory below the compactification scale $R^{-1}$, where $R$ is the
distance between the visible and hidden branes.
If the scale of new physics is $M$ (\eg\ the string scale),
then for $R \gg M^{-1}$
the propagator of a massive field (\eg\ and excited string state)
connecting the visible and hidden
branes is suppressed by the Yukawa factor $e^{-MR} \ll 1$.
(Since the suppression factor is exponential, $R \gsim \hbox{\rm few}
\times M^{-1}$ is sufficient in practice.)
Therefore, operators like \Eq{scalarMPopagain} are not generated
by the exchange of massive states.
This leaves only the effect of fields that are light compared to the
compactification scale.
Only supergravity \emph{must} propagate in the bulk, so
the minimalthe minimal light
fields in the bulk are the minimal 5D SUGRA multiplet.
It was shown in \Ref{LS} that this does not generate terms of
the form \Eq{scalarMPopagain}.
For details, see \Refs{RS0,LS}.
In this setup, the SUSY breaking sector is `more hidden' than in conventional
hidden sector models, and is sometimes referred to as a `sequestered
sector.'

Another way to make this natural is to replace the extra dimension
in the setup above with a conformal field theory via the AdS/CFT
correspondence.
For a discussion of this `conformal sequestering,'
see \Ref{LSConformalSequestering}.

% -----------------------------------------------------------------------------
\section{Gaugino Mediation}
The final model we will mention (very briefly) is gaugino mediation.
Like anomaly mediation, this can be motivated by an extra-dimensional
setup.
This time we assume that the standard-model gauge fields propagate
in the bulk, while the matter fields are localized on the visible
brane.
In this case, the gauginos can get a mass from contact terms on
the hidden brane of the form
\beq
\De\scr{L} \sim \myint d^2\th\, \frac{X}{M} W^\al W_\al + \hc,
\eeq
where $X$ is the hidden sector field that breaks SUSY.
In this type of model, the gaugino gets a mass at tree level, while
the visible matter fields get a mass only at one-loop order.
This means that the gaugino masses are much larger than scalar masses
at the compactification scale, but the RG between this scale and the
weak scale generates scalar masses of order the gaugino masses.
This scenario is called `gaugino mediated SUSY breaking' for obvious
reasons.
The original papers are \Refs{gauginomed}.

Note that gaugino mediation shares the fine-tuning problem of gauge-
and anomaly-mediation.
In a GUT model, we expect that the gaugino masses are unified at
the GUT scale:
\beq
M_1(M_{\rm GUT}) \simeq M_2(M_{\rm GUT}) \simeq M_3(M_{\rm GUT}).
\eeq
(Even if there are GUT breaking splittings, we expect $M_1 \sim M_2
\sim M_3$, which is sufficient for our argument.)
Since the quantity $M_i / g_i^2$ is RG invariant at one loop, at the
weak scale we have
\beq
\frac{M_1}{g_1^2} \simeq \frac{M_2}{g_2^2} \simeq \frac{M_3}{g_3^2}.
\eeq
The scalar masses are generated from the RG equation
\beq
\frac{d m^2}{d t} = -\frac{c}{2\pi^2} g^2 m_\la^2.
\eeq
Using the fact that $m_\la^2 / g^2$ is RG invariant, we have the
solution
\beq
m^2(\mu) = \frac{2c}{b} \left[ g^4(\mu) - g^4(M_{\rm GUT}) \right]
\left( \frac{m_\la}{g^2} \right)^2,
\eeq
where we have assumed that $m^2(M_{\rm GUT}) \ll m^2(\mu)$.
We therefore have
\beq
\frac{m_{\tilde q}^2}{m_{\tilde{e}_R}^2}
\sim \frac{N_{\rm c} g_3^4}{g_1^4}
\eeq
just as in gauge mediation and anomaly mediation.

% -----------------------------------------------------------------------------
\section{No Conclusion}
There is much more to say, but I will stop here.
I hope that I have introduced some of the problems and issues with
SUSY breaking, as well as introducing some ideas that may point
in the right direction.
I hope that some of the readers will be inspired by these lectures
to go beyond them.

\end{document}